\documentclass{article}
\usepackage{xcolor}
\usepackage{bbding} 

\usepackage{jheppub}
\usepackage[utf8]{inputenc}
\usepackage{amsmath}
\usepackage{amssymb}
\usepackage{amsfonts}
\usepackage{amsthm}
\usepackage{appendix}
\makeatletter
\DeclareRobustCommand*{\bfseries}{%
   \not@math@alphabet\bfseries\mathbf
   \fontseries\bfdefault\selectfont
   \boldmath
}
\makeatother

\def\bO{\bar 0}
\def\b1{\bar 1}
\def\lieg{\mathfrak{g}}
\def\lieh{\mathfrak{h}}
\def\liem{\mathfrak{m}}
\def\liep{\mathfrak{p}}
\def\liek{\mathfrak{k}}
\def\id{\textit{id}}
\def\Spin{\textit{Spin}}

\title{The Superconformal \XSolidBrush ing Equation}

\preprint{DESY 20-096}

\author{Ilija Buri\' c$^1$,}
\author{Volker Schomerus$^1$}
\author{and Evgeny Sobko$^2$}
\affiliation{$^1$DESY, Notkestra\ss e 85, D-22607 Hamburg, Germany}
\affiliation{$^2$University of Southampton, Highfield, Southampton,
SO17 1BJ, United Kingdom}
\date{May 2020}

\abstract{Crossing symmetry provides a powerful tool to access the
non-perturbative dynamics of conformal and superconformal field theories.
Here we develop the mathematical formalism that allows to construct the
crossing equations for arbitrary four-point functions in theories with
superconformal symmetry of type I, including all superconformal
field theories in $d=4$ dimensions. Our advance relies on a supergroup
theoretic construction of tensor structures that generalizes an approach
which was put forward in \cite{Buric:2019dfk} for bosonic theories.
When combined with our recent construction of the relevant superblocks, we are
able to derive the crossing symmetry constraint in particular for four-point
functions of arbitrary long multiplets in all 4-dimensional superconformal
field theories.}

\begin{document}
\addtolength{\baselineskip}{2mm}
\maketitle

\section{Introduction}

Conformal field theories describe very special points in the space of quantum field theories
that seem to provide unique views into non-perturbative dynamics through a variety of rather
complementary techniques, such as holography, integrability, localization and the conformal
bootstrap. One of the principal analytical tools for conformal field theory are conformal
partial wave (or block) expansions that were proposed early on in \cite{Ferrara:1973vz}.
The role they play in the study of models with conformal symmetry is very similar to the
role of Fourier analysis in systems with translational symmetry. While conformal blocks
are entirely determined by kinematics, they allow to separate very neatly the dynamical
meat of a theory from its kinematical bones. For example, an $N$-point function of local
operators in a conformal field theory can be a very complicated object. If expanded in
conformal blocks, however, the coefficients factorize into a set of three-point couplings,
i.e.\ most of the complicated dependence on the insertion points resides in the kinematical
skeleton of a conformal field theory. This is the reason conformal blocks expansions
are so important.

Conformal blocks for four-point functions of local operators in bosonic conformal field
theories are relatively well studied by now, see e.g. \cite{Dolan:2000ut,Dolan:2003hv,
Dolan:2011dv,Costa:2011dw,SimmonsDuffin:2012uy,Penedones:2015aga,Hogervorst:2013sma,
Echeverri:2016dun,Schomerus:2016epl,Karateev:2017jgd,Isachenkov:2017qgn,Dyer:2017zef,Erramilli:2019njx,Fortin:2019fvx,Fortin:2019dnq,
Fortin:2020ncr} and references therein. On the other hand, while we know of many
examples of such theories in $d=3$ dimensions, most conformal field theories in
$d\geq 4$ seem to possess supersymmetry. The enhancement from conformal to superconformal
symmetry should lead to simplifications, at least once the kinematical aspects are well
under control. This, however, is not yet the case. In fact, while four-point blocks of
half-BPS operators or the superprimary components of more general supermultiplets have
been constructed and applied, see e.g. \cite{Dolan:2001tt,Dolan:2004mu,Nirschl:2004pa,
Poland:2010wg,Fortin:2011nq,Fitzpatrick:2014oza,Khandker:2014mpa,Bobev:2015jxa,Bissi:2015qoa,
Doobary:2015gia,Lemos:2015awa,Liendo:2016ymz,Lemos:2016xke,Chang:2017xmr,Bobev:2017jhk,
Liendo:2018ukf,Berkooz:2014yda,Li:2016chh,Li:2017ddj,Gimenez-Grau:2019hez}, relatively little
is actually known about blocks and block expansions for more generic external multiplets
that span long(er) representations of the superconformal algebra. On the other hand it
has been shown in \cite{Cornagliotto:2017dup} that the bootstrap with long multiplets
is significantly more constraining on CFT data than the bootstrap with e.g. external
BPS operators, see also \cite{Kos:2018glc}. This provides strong motivation
to investigate blocks and crossing symmetry for long multiplets, which is the main
goal of our work.
\medskip

In order to explain the main results of this paper, let us briefly review a few basic
facts about conformal partial wave expansions in bosonic conformal field theories. We
start from some four-point correlator $G(x_i)$ with its full dependence on the insertion
points $x_i$ of the fields. As is well known, conformal symmetry implies that $G(x_i)$
is fully determined by a some function of the two cross ratios $u,v$ one can form from
four points in $\mathbb{R}^d$. More precisely, it is possible to write the correlation
function $G$ as
\begin{equation}
G(x_i) = \Omega(x_i) g(u,v) \ .
\end{equation}
We stress that such a behavior is not restricted to scalar correlation functions. If
the fields carry spin, then $G$ takes values in the space of polarizations of the
four fields. The function $g$, on the other hand takes values in the space of four-point
tensor structures whose dimension is smaller than that of the space of polarizations,
in general, at least for $d > 3$. Hence, one should think of $\Omega$ as a rectangular
matrix. We shall refer to such a matrix valued function $\Omega$ of the insertion points
as four-point \textit{tensor factor}. In some sense to become clear below it combines all
four-point tensor structures into one single object $\Omega$. Many authors have studied
tensor structures for spinning four-point functions in conformal field theories, see
e.g. \cite{Osborn:1993cr,Costa:2011mg,Costa:2011dw,Kravchuk:2016qvl,Cuomo:2017wme,Karateev:2018oml,
Karateev:2019pvw}.

The tensor factor $\Omega(x_i)$ is restricted but not determined by conformal symmetry.
In fact, there is some obvious `gauge' freedom that is associated with matrix-valued
functions $\zeta(u,v)$ one can move back and forth between the tensor factor $\Omega$ and the function $g(u,v)$,
i.e.\ the gauge symmetry acts as $(\Omega,g) \rightarrow (\Omega \zeta^{-1}, \zeta g)$. The
function $g$ of the cross ratios may be expanded in terms of conformal partial waves which,
after the influential work of Dolan and Osborn \cite{Dolan:2000ut,Dolan:2003hv}, are characterised
as eigenfunctions of the so-called Casimir differential equations. The form of these equations,
however, depends on the gauge choice that is made when splitting $G$ into $\Omega$ and $g$. For
four-point functions of identical scalar fields of weight $\Delta_0$, for example, Dolan and Osborn
chose $\Omega_s = x_{12}^{-2\Delta_0} x^{-2\Delta_0}_{34}$. Note that this factor $\Omega =
\Omega_s$ also depends on a split of the four points into two sets of two, a choice usually
referred to as a channel. Here we have displayed the factor $\Omega$ for the so-called
$s$-channel. The $t$-channel is obtained by exchanging the fields inserted at $x_2$ and
$x_4$. With their pick of $\Omega_s$, Dolan and Osborn worked out the associated
Casimir differential equation for the function $g_s$ and similarly for $g_t$. Solutions
of these Casimir equations provided them with a set of blocks $g_{s/t}^{\Delta, l}(u,v)$
in which one can then expand $g_s$ and $g_t$
\begin{equation}
G(x_i) = \Omega_s(x_i) \sum p_{\Delta,l} g^{\Delta,l}_s (u,v) =
\Omega_t(x_i) \sum p_{\Delta,l} g^{\Delta,l}_t (u,v) \ .
\end{equation}
The equality between the first and second sum is the famous crossing symmetry equation.
An important observation is that writing this equation does actually not require a complete
knowledge of the tensor factors. It is sufficient to know the ratio of the $s$- and
$t$-channel $\Omega$
$$
M(u,v) = \Omega_t^{-1}(x_i) \Omega_s (x_i) =  \left(\frac{v}{u}\right)^{\Delta_0}\ ,
$$
which is a function of the two cross ratios only. We call this important object $M$
the \textit{crossing factor} $M$. In the case of spinning fields the crossing factor
becomes a matrix. This ratio of $s$- and $t$-channel tensor factors is not to be
confused with the crossing or fusing matrix of the conformal group. While the
crossing factor relates the $s$- and $t$-channel tensor factors, the crossing matrix
relates the conformal blocks in the two channels by providing the expansion
coefficients of $s$-channel blocks in terms of $t$-channel ones, see
\cite{Liu:2018jhs,Sleight:2018ryu,Chen:2019gka} for some recent discussion
in the context of higher dimensional conformal field theory.

In \cite{Isachenkov:2016gim} it was noticed that scalar four-point functions $G$ admit an different
gauge choice for the factor $\Omega$ such that the associated Casimir equations take the
form of an eigenvalue equation for an integrable 2-particle Hamiltonian of Calogero-Sutherland
type. This was later explained in \cite{Schomerus:2016epl,Schomerus:2017eny} through harmonic
analysis on the conformal group and then extended to fields with spin in which case the quantum
mechanical potential becomes matrix valued. For spinning fields, the tensor structures of such a
Calogero-Sutherland gauge were constructed recently in \cite{Buric:2019dfk}. The goal of our work
is to extend all this to the case of superconformal symmetry. In \cite{Buric:2019rms} we have
constructed the Casimir equations for superconformal symmetries of type I. The form of these equations
allows us to compute superblocks systematically as finite sums of spinning bosonic blocks.
What was missing up to now is the construction of  the associated tensor structures and in particular
the crossing factor $M$. Below we  fill this gap and construct both the tensor structures
and the crossing factor for all superconformal algebras of type I. Explicit formulas
for the crossing factors in 4-dimensional superconformal algebras will be given in our
forthcoming paper \cite{N1D4_paper}. Early work on tensor structures for four-point
correlators of superconformal field theories includes \cite{Park:1997bq,Park:1999pd,Osborn:1998qu,Heslop:2002hp,Heslop:2004du,Nirschl:2004pa}.
\medskip

Let us now describe the plan of this work in more detail. The next section contains some
basic background material on superconformal algebras, where we introduce the notion of
superspace and discuss the infinitesimal and global action of the conformal symmetry
thereon. Special attention will be paid to the action of the so-called Weyl inversion,
which plays an important role in later sections. Section 3 contains the first new
result of this work. There we construct a special family $g(x_i)$ of supergroup elements
that depend on the insertion points of the fields along with a matrix realization that
uniquely encodes the correlation function $G(x_i)$. This generalizes a similar formula
for bosonic conformal field theories in \cite{Buric:2019dfk} to the supersymmetric setup.
In section 4 we begin to specialize the discussion to superconformal algebras of type I,
i.e.\ to cases in which the R-symmetry group contains an abelian factor $U(1)$. After
introducing super Cartan coordinates through a particular KAK factorization of the
superconformal group we can construct the tensor factors $\Omega$ for any choice
of spins and any channel through an elegant group theoretical construction. This
then allows us to build the crossing factor $M$ as a quotient of $s$- and $t$-channel
tensor factors and prove its conformal invariance explicitly. Our main result, which is
stated in eqs.\ \eqref{eq:crossingmatdef}, \eqref{eq:crossingmatrix}, expresses the crossing
factor $M$ through representation matrices of some particular family of elements of
the group $K$ that is generated by dilations, rotations and R-symmetry transformations.
All constructions in section 2-4 are illustrated at the example of $\mathfrak{g} =
\mathfrak{sl}(2|1)$ of the $\mathcal{N} = 2$ superconformal algebra in $d=1$ dimensions.
Let us also note that our discussion includes the purely bosonic case $\mathcal{N}=0$
for which the crossing factor was not constructed previously beyond a few special spin
assignments.  As a corollary to our discussion we state the crossing factor for arbitrary
spinning four-point functions in 3-dimensional conformal field theories. For all other
higher dimensional examples, bosonic as well as supersymmetric, our results for the
crossing factor are stated in the form of a precise easy-to-follow algorithm. In
order to obtain ready-to-use formulas one needs to input some classical results
from the group theory of rotations $SO(d)$. We will discuss this for certain
mixed correlators in $\mathcal{N}=1$ superconformal theories in an accompanying
work \cite{N1D4_paper}.

\section{Superspace and Superconformal Symmetry}

In order to state and prove our main results we need some background on supergroups,
superspaces and the action of superconformal symmetry thereon. Here we want to review
these concepts and at the same time introduce a mathematical language that is appropriate
for our subsequent discussion. In particular, we  recall the notion of superspace
in the second subsection and explain how one constructs an infinitesimal action of the
superconformal algebra thereon. This action is lifted to global transformations in the
third subsection, with some special focus on the so-called Weyl inversion, a close
relative of the conformal inversion which is guaranteed to exist in any superconformal
field theory. For more mathematical minded readers we have incorporated a more abstract
and introductory subsection on the concept of supergroups. While this helps to make
equations in subsequent subsections mathematically rigorous, readers who feel familiar
with supergroups and superspaces are encouraged to skip the first subsection, at least
upon first reading.

\subsection{Some basics on superalgebras and supergroups}

In this subsection we  introduce some very basic notions and notations concerning
supergroups. Our conventions agree with \cite{Kostant:1975qe,Leites:1980rna,Wess:1992cp}.
Let $\mathfrak{h}$ be some Lie superalgebra, i.e. a graded vector space $\mathfrak{h} =
\mathfrak{h}_{\bO} \oplus \mathfrak{h}_{\b1}$ with a graded Lie bracket.
We denote the latter by $[. , . ]_\pm$. The associated \textit{universal enveloping
algebra} $U(\mathfrak{h})$ is the graded associative algebra generated by elements $X  \in
\mathfrak{h}$, with relations such that graded commutators are given by the Lie bracket.
In a slight abuse of notations we shall denote the graded commutators in the universal
enveloping algebra by $[.,.]_\pm$ as well.

The universal enveloping algebra comes equipped with a co-product $\Delta$, i.e. with
a homomorphism
$$ \Delta: U(\lieh) \rightarrow U(\lieh) \otimes U(\lieh)\ . $$
Here, the tensor product is to be understood in the graded sense, i.e. elements are
multiplied as
$$ (a_1 \otimes b_1) \cdot (a_2 \otimes b_2) = (-1)^{|a_2||b_1|} a_1 a_2 \otimes
b_1 b_2  \ ,$$
where $|a|=0$ if $a$ is even and $|a|=1$ if $a$ is odd, as usual. On the generating
elements $X \in \lieh \subset U(\lieh)$, the co-product is given by
\begin{equation}
\Delta(X) = X \otimes 1 + 1 \otimes X \ .
\end{equation}
From here one can extend $\Delta$ uniquely to the entire universal enveloping algebra
as a homomorphism of graded algebras. The co-product is the algebraic structure that
allows us to build tensor products of any two representations of the Lie superalgbra
$\mathfrak{h}$ or its universal envelop $U(\mathfrak{h})$.
\medskip

Let us now turn to another algebra that we can associate to $\lieh$, namely the so-called
\textit{structure algebra} $\mathcal{F}(\mathfrak{h})$. By definition, $\mathcal{F}$ is a
graded commutative algebra whose generators $x_A$ are associated to the basis elements $X^A$
of the Lie superalgebra $\lieh$. The elements $x_A$ possess the same degree $|x_A|= |A|$
as the generators $X^A$, i.e. $x_A$ is an ordinary bosonic variable if $X^A$ is even while
$x_A$ is a Grassmann variable in case $X^A$ is odd. From the construction we have sketched
here it is evident that $\mathcal{F}$ can be thought of as the \textit{algebra of functions}
on the supergroup associated with $\lieh$ which is generated here from set of coordinate
functions, one for each element of the Lie superalgebra.
\smallskip

The two algebras we have associated to $\lieh$ up to now are actually closely related.
In the case of bosonic groups, the generators $X$ of the Lie algebra give rise to (right)
invariant vector fields that act on functions as some first order differential operators.
These differential operators $\mathcal{R}_X$ can be multiplied and added and thereby
provide an action of elements $a$ in the universal enveloping algebra $U(\lieh)$ through
differential operators $\mathcal{R}_a$ of higher order. One may combine the application
of any such differential operator to a function on the group with the evaluation at the
group unit $e$ to obtain a map that assigns as number
\begin{equation} \label{eq:duality}
\mathcal{R}_a(f)(e) = (a,f) = f(a) \in \mathbb{C}
\end{equation}
to a pair of an element  $ a \in U(\lieh)$ and a (complex valued) function $f$ on the
group. In other words, elements of $U(\lieh)$ give linear functionals of the algebra of
functions or structure algebra $\mathcal{F}(\lieh)$ and vice versa. In this form, the
statement remains true for Lie superalgebras and is often expressed by saying that
$\mathcal{F}(\lieh)$ and $U(\lieh)$ are dual to each other, see also \cite{Sternberg:1975}
for a nice discussion of this point.
\medskip

Equipped with these two algebraic structures, namely the universal enveloping algebra
$U(\lieh)$ and the structure algebra $\mathcal{F}(\lieh)$, we want to introduce the concept
of \textit{supergroup elements} $h$. Let us first give a formal definition according to
which $h$ is an even element of the graded tensor product $U(\lieh) \otimes \mathcal{F}(\lieh)$
that satisfies
\begin{equation} \label{eq:Deltah}
(\Delta \otimes \id) h = \ \stackrel{1}{h}\  \stackrel{2}{h} \ .
\end{equation}
Here, the application of the co-product $\Delta$ to the first tensor factor of $h$
produces an element in $U(\lieh) \otimes U(\lieh) \otimes \mathcal{F}(\lieh)$. The
factors on the right hand side are elements in the same threefold tensor product.
More concretely, $\stackrel{2}{h}$ is the element $1 \otimes h$ with trivial entry
in the first tensor factor. Similarly $\stackrel{1}{h}$ denotes the element $h$ with
trivial entry in the second tensor factor.

The element $h$ is not uniquely characterized by these properties, but we do not need
to be more specific. It might be helpful to think of $h$ as the object $h = \exp (x_A
X^A)$. The element $x_A X^A$ in the exponent is even and upon expansion of the
exponential provides us with an even element in the graded tensor product $U(\lieh)
\otimes \mathcal{F}(\lieh)$. In order to construct this element one moves all the
elements $x_A$ of the structure algebra to the right of the superalgebra generators
$X^B$ using
$$ x_A X^B = (-1)^{|A| |B|} X_B x_A \  ,$$
which implements our convention to consider the graded tensor product of $U(\lieh)$
and $\mathcal{F}(\lieh)$ rather than the ordinary one. After the reordering we indeed
obtain an infinite sum of products between elements in the universal enveloping algebra
$U(\lieh)$ with elements of the structure algebra $\mathcal{F} (\lieh)$. If we apply
the co-product in the universal enveloping algebra we formally
obtain
\begin{equation}
(\Delta \otimes \id) h = e^{x_A (X^A \otimes 1 + 1 \otimes X^A)} = e^{x_A (X^A \otimes 1)}
e^{x_A (1 \otimes X^A)} =\  \stackrel{1}{h}\  \stackrel{2}{h} \ .
\end{equation}
In writing the single exponential as a product of exponentials we used the fact that
the exponent if an even object so that $x_A (X^A \otimes 1)$ commutes with $x_A (1
\otimes X^A)$. In conclusion, we have constructed an object $h$ with the properties
we demanded in the previous paragraph, at least formally. In physics, it is
customary to evaluate $h$ in some representation $\pi$ of the Lie superalgebra $\lieh$
or, equivalently, its universal enveloping algebra. Thereby one obtains a finite
dimensional supermatrix $h^\pi = (\pi \otimes \id) h$ with entries from the structure
algebra $\mathcal{F}$. In the following we  often use the symbol $h$ for such a
matrix $h$ rather than the universal element $h \in U(\lieh) \otimes \mathcal{F}(\lieh)$.
\medskip

What we have explained so far actually suffices as background for most of our
discussion below, except for the construction of an infinitesimal action of the
conformal superalgebra on superspace in the next subsection. To obtain explicit
formulas for the first order derivative operators $\mathcal{R}_X$  that are
associated with the elements $X \in \lieh$ let us first extend the structure
algebra $\mathcal{F}(\lieh)$ of ``functions on the supergroup'' to a differentially
graded algebra $d\mathcal{F}(\lieh)$ of ``differential forms on the supergroup''.
The latter is a bi-graded commutative algebra generated by elements $x_A$ and
$dx_A$, with a second grading associated to the form degree.
On the algebra $d\mathcal{F}(\lieh)$ we can define a differential $d$ that
squares to zero $d^2 = 0$ and satisfies the graded Leibniz rule
$$ d (f \wedge g) = df \wedge g + (-1)^{\textit{deg}(f)} f \wedge dg \ . $$
Here $\textit{deg}(f)$ denotes the form degree of $f$. Let us stress that there is no
additional sign associated with the $\mathbb{Z}_2$ grading that distinguishes between even
(bosonic) and odd (fermionic) elements. This means that $d$ is treated as an even object.
Hence, for a given $A$, $x_A$ and $dx_A$ possess the same degree, i.e.\ $dx_A$ is even
[odd] in case $x_A$ is even [odd].
\medskip

Since the structure algebra $\mathcal{F}(\lieh)$ is contained in the larger differentially
graded algebra $d\mathcal{F}(\lieh)$ we can also think of the supergroup element $h \in
U(\lieh) \otimes \mathcal{F}(\lieh)$ as an element of the differential graded algebra
$U(\lieh) \otimes d\mathcal{F}(\lieh)$ with the additional rule that $dX^A= X^Ad$, i.e.\
we consider the generators $X^A$ of the Lie superalgebra as constants and the differential
$d$ as even. Now it makes sense to consider the Maurer-Cartan form
\begin{equation}
dh h^{-1} \in U(\lieh) \otimes d\mathcal{F}(\lieh) \ .
\end{equation}
If we apply the differential to the equation \eqref{eq:Deltah} that characterizes
$h$ we obtain
\begin{equation}
\Delta(dh h^{-1}) = \left(\stackrel{\phantom{0}}{d}\stackrel{1}{h} \  \stackrel{2}{h} +
\stackrel{1}{h} \ \stackrel{\phantom{0}}{d}\stackrel{2}{h}\right)\  \stackrel{2}{h}\!^{-1} \
\stackrel{1}{h}\!^{-1}
= \ \stackrel{\phantom{0}}{d}\stackrel{1}{h} \ \stackrel{1}{h}\!^{-1} +
 \ \stackrel{\phantom{0}}{d}\stackrel{2}{h} \ \stackrel{2}{h}\! ^{-1}\ .
\end{equation}
We conclude that the Maurer-Cartan form takes values in the Lie superalgebra $\lieh
\subset U(\lieh)$, as it is the case for usual bosonic Lie groups. Consequently, it
may be expanded as
\begin{equation}
dh h^{-1} = dx_A C_{AB} X^B\quad \textit{where} \quad C_{AB} \in
\mathcal{F}(\lieh)\ .
\end{equation}
The matrix elements $C_{AB}$ possess degree $|A|+|B|$, i.e.\ they are even elements
of the structure algebra if $|A|=|B|$ and odd otherwise. We also stress that the elements
$C_{AB}$ depend on the choice of the supergroup element $h$. One of the main uses of the
matrix elements $C_{AB}$ is to construct the right-invariant vector fields, i.e. an action
of the Lie superalgebra $\lieh$ through first order differential operators acting on the
structure algebra $\mathcal{F}(\lieh)$. These vector fields are given by
\begin{equation} \label{eq:RHA}
    \mathcal{R}_{X^A} = \mathcal{R}_A :=
    \mathcal{C}^G_{AB}\partial_B, \nonumber
\end{equation}
where $\mathcal{C} = C^{-1}$ denotes the inverse of $C$ and $\partial_B$ is the (graded)
derivative with respect to the coordinate $x_B$. Its action on an arbitrary function $f
\in \mathcal{F}(\lieh)$ can be read off from $df = dx_B (\partial_B f)$. In particular,
when acting on the individual coordinate functions, $x_A$ is obeys $(\partial_B x_A) =
\delta_{A,B}$. The action of partial derivatives on products of functions satisfies the
graded Leibniz rule which implies that
\begin{equation}
\partial_B x_A = (\partial_B x_A) + (-1)^{|A||B|} x_A \partial_B = \delta_{A,B} +
(-1)^{|A||B|} x_A \partial_B \ .
\end{equation}
Since we have assumed that the differential $d$ acts trivially on the generators
$X^A$ of the universal enveloping algebra, i.e. $(dX^A) = 0$ we conclude that
$\partial_\beta X^A = 0$, i.e.\ the generators $X^A$  are constant objects on
the supergroup statisfying
\begin{equation}
\partial_B X^A = (-1)^{|A||B|} X^A \partial_B \ \ .
\end{equation}
With this list of properties of the partial derivatives we conclude our construction
of the right invariant vector fields \eqref{eq:RHA} and thereby our short mathematical
review of superalgebras and the theory of supergroups. The formulation we have introduced
here is well adapted to our needs below and also paves the way for some interesting
extensions, see the concluding section.

\subsection{Superspace and the infinitesimal action of superconformal symmetry}

This subsection serves two purposes. On the one hand we need to introduce the notion
of superspace that is one of the crucial ingredients throughout the rest of the paper.
In addition we shall also construct an action of the superconformal algebra $\lieg$
through ``first differential operators'' on superspace. This infinitesimal action of
the superconformal symmetry on superspace will play only a minor role below since
most of our analysis is based on global transformations.

To set up notations let us denote the superconformal algebra by $\mathfrak{g}$. Its
bosonic subalgebra $\lieg_{\bO}$ consists of $d$-dimensional conformal transformations
in $\mathfrak{so}(1,d+1)$ as well as R-symmetry transformations in some Lie algebra
$\mathfrak{u}$. To define superspace we pick some decomposition
\begin{equation}\label{eq:decomposition}
    \mathfrak{g} = \mathfrak{m} \oplus \mathfrak{p}  \nonumber
\end{equation}
of $\mathfrak{g}$ into two Lie subalgebras $\mathfrak{p}$ and $\mathfrak{m}$. The standard
choice would be to define $\mathfrak{p}$ as the span of all elements in $\mathfrak{g}$ that
lower the eigenvalue of the dilation generator $D \in \mathfrak{g}_{\bO}$, i.e.
$$ \mathfrak{p} :=  \lieg_{\leq 0} =
\textit{span}\left(\, X \in \mathfrak{g}\, | \, [D,X] = \alpha X
\, , \, \alpha \leq 0 \right)\ . $$
For this choice, $\mathfrak{m}$ then consists of generators $P$ of translations and
the supercharges $Q$. We shall briefly comment on other choices below. We also choose
a basis $X^A$ of elements in $\lieg$ that is compatible with the decomposition
\eqref{eq:decomposition}. Elements $X^A$ that lie in the subspace $\liem$ will be
labeled by lower case Latin indices while those that lie in the complement $\liep$
carry Greek indices.

The decomposition of the Lie superalgebra $\lieg$ into $\liem$ and $\liep$ determines
a decomposition of the corresponding universal enveloping algebra $U(\mathfrak{g})=
U(\mathfrak{m})\otimes U(\mathfrak{p})$ as well as of the structure algebra
$\mathcal{F}(\lieg)=\mathcal{F}(\liem)\otimes\mathcal{F}(\liep)$. Recall that the
structure algebras $\mathcal{F}(\liem)$ and $\mathcal{F}(\liep)$  are generated by
the coordinates $x_a$ and $x_\alpha$, respectively, with $x_a$ and $x_\alpha$ being
Grassmann variables if the corresponding elements $X^a$ and $X^\alpha$ are fermionic
generators of the Lie superalgebra. The structure algebra $\mathcal{F}(\liem)$ is what
is referred to a \textit{superspace} $\mathcal{M} = \mathcal{F}(\liem)$. Loosely
speaking one may think of it as the algebra of ``functions on the supergroup $M$'',
though we have not defined what we mean by a supergroup and do not intend to do so.
\medskip

Now that we know what superspace is let us construct an infinitesimal action of
the superconformal symmetry thereon. Here we shall closely follow the general
constructions we outlined in the previous subsection and introduce supergroup
elements $m=m(x_a)$ and $p=p(x_\alpha)$. In case of $m$ we work with the
following standard choice
\begin{equation}
m(x_a) = e^{x_a X^a}  .
\end{equation}
The infinitesimal action of the conformal algebra on the coordinates $x_a$ of our
superspace descends from the left-regular action of $\lieg$ and thus can be
computed from the Maurer-Cartan form,
\begin{equation}
    dg g^{-1} = dx_A C^{G}_{AB}X^B\ .
\end{equation}
In computing the Maurer-Cartan form for $\lieg$ it is usual to relate it to the
Maurer-Cartan forms that are associated with $\mathfrak{m}$ and $\mathfrak{p}$
\begin{equation}
    dm m^{-1} = dx_a C^M_{ab} X^b \quad , \quad
    dp p^{-1} = dx_\alpha C^{P}_{\alpha\beta} X^\beta\ .\nonumber
\end{equation}
With our choice $g=mp$ of the supergroup element $g$ as a product of the two
elements $m$ and $p$ it follows that
\begin{align}
    & dg g^{-1} = dx_A\partial_A(m p) (m p)^{-1} = dx_a (\partial_a m) m^{-1} +
    dx_\alpha m (\partial_\alpha p) p^{-1} m^{-1} =\nonumber\\[2mm]
    & = dx_a C^M_{ab} X^b + dx_\alpha m C_{\alpha\beta}X^\beta m^{-1} =
    dx_a C^M_{ab} X^b + dx_\alpha C^P_{\alpha\beta}\Big((M_1)_{\beta a} X^a +
    (M_2)_{\beta\gamma}X^\gamma\Big) \ . \label{MC-form}
\end{align}
The last equality defines the two matrices $M_{1,2}$,
\begin{equation}
    m X^\beta m^{-1} = (M_1)_{\beta a} X^a + (M_2)_{\beta\gamma} X^\gamma\ .
\end{equation}
From the equation $(\ref{MC-form})$ we can read off the coefficients $C^G_{AB}$
of the Maurer-Cartan form for $\lieg$. The inverse $\mathcal{C}^G$ of this matrix
is easily seen to take the form
\begin{equation}
    \mathcal{C}^G = \begin{pmatrix}
    \mathcal{C}^M & 0 \\
    -M_2^{-1}M_1\mathcal{C}^M & M_2^{-1} \mathcal{C}^P
    \end{pmatrix} \ , \nonumber
\end{equation}
where the first row/column corresponds to direction in $\liem$ while the second
row/column collects all the directions in $\liep$. As stated before, the matrix
$\mathcal{C}^G$ provides us with the right-invariant vector fields \eqref{eq:RHA}
on the conformal supergroup. To project these operators to the superspace one
simply sets $\partial_\alpha=0$,
\begin{equation}\label{eq:resultRM}
    \mathcal{R}^{(M)} = \begin{pmatrix}
    \mathcal{C}^M & 0\\
    -M_2^{-1}M_1\mathcal{C}^M & M_2^{-1} \mathcal{C}^P
    \end{pmatrix}
    \begin{pmatrix} \partial\\ 0 \end{pmatrix} =
    \begin{pmatrix} \mathcal{C}^M_{ab}\partial_b\\
    -(M_2^{-1}M_1\mathcal{C}^M)_{\alpha b}\partial_b
    \end{pmatrix}\ . \nonumber
\end{equation}
This is the main result of this subsection. As mentioned above, the differential
operators on superspace depend on $C^M$ and hence on the choice of the supergroup
element $m$. The choice of the supergroup element $p$, on the other hand, is
irrelevant since the coefficients $C^P$ of the Maurer-Cartan form $dp p^{-1}$
dropped out in the last step when we set all derivatives $\partial_\alpha$ to
zero.
\smallskip

Our result \eqref{eq:resultRM} applies to all decompositions of $\lieg$ into two Lie
subalgebras $\liem$ and $\liep$. As we pointed out in the first paragraph, the standard
choice is to take $\liep$ to contain generators that do not increase the conformal weight.
In that case, the structure algebra $\mathcal{M} = \mathcal{F}(\liem)$ is called the standard
superspace. If the superconformal algebra $\mathfrak{g}$ is of type I, however, there
exist other natural choices to which the constructions of this subsection apply. In a
type I superalgebra the R-symmetry contains a $U(1)$ subalgebra which commutes
with all bosonic generators but assigns the fermionic ones a non-trivial
R-charge $\pm 1$. As usual, we can decompose the Lie superalgebra $\lieg =
\lieg_{\leq 0} \oplus \lieg_{> 0}$ by splitting off those generators in
$\lieg_{>0}$ that strictly increase the conformal weight. These consist
of supercharges $Q$ and generators of translations. In a type I superalgebra
we can now split the space $\mathfrak{q}$ or supercharges $Q$ according to
the sign of their $U(1)$ R-charge as $\mathfrak{q} = \mathfrak{q_+} \oplus
\mathfrak{q}_-$. With this in mind we can introduce two new decompositions
$\lieg = \liem_\pm \oplus \liep_\pm$ of the superconformal algebra where
 \begin{equation}
    \mathfrak{p}_\pm = \mathfrak{g}_{\leq 0} \oplus \mathfrak{q}_\pm \ , \quad
    \mathfrak{m}_\pm = \mathfrak{g}_1\oplus\mathfrak{q}_\mp = \mathfrak{g}/
    \mathfrak{p}_\pm \ . \nonumber
\end{equation}
From the properties of type I Lie superalgebras, one may easily show that both
$\mathfrak{p}_\pm$ and $\mathfrak{m}_\pm$ are subalgebras of $\mathfrak{g}$.
The associated superspaces $\mathcal{M}_\pm = \mathcal{F}(\liem_\pm)$ are
called the chiral and anti-chiral superspace, respectively.
\bigskip

\noindent
{\bf Example:} As an example, let us illustrate the construction of superspace and the differential
operators in the case of the 1-dimensional $\mathcal{N}=2$ superconformal algebra $\lieg=\mathfrak{sl}(2|1)$.
The smallest faithful representation of $\mathfrak{g}$ is 3-dimensional. We may choose the generators
as
\begin{equation} \label{eq:bosrep}
    D = \begin{pmatrix}
    1/2 & 0 & 0\\
    0 & -1/2 & 0\\
    0 & 0 & 0
    \end{pmatrix},\ P = \begin{pmatrix}
    0 & 1 & 0\\
    0 & 0 & 0\\
    0 & 0 & 0
    \end{pmatrix},\ K = \begin{pmatrix}
    0 & 0 & 0\\
    1 & 0 & 0\\
    0 & 0 & 0
    \end{pmatrix},\ R = \begin{pmatrix}
    -1 & 0 & 0\\
    0 & -1 & 0\\
    0 & 0 & -2
    \end{pmatrix},\nonumber
\end{equation}
for the four bosonic generators and
\begin{equation} \label{eq:fermrep}
    Q_- = \begin{pmatrix}
    0 & 0 & 0\\
    0 & 0 & 0\\
    0 & 1 & 0
    \end{pmatrix},\ Q_+ = \begin{pmatrix}
    0 & 0 & 1\\
    0 & 0 & 0\\
    0 & 0 & 0
    \end{pmatrix},\ S_- = \begin{pmatrix}
    0 & 0 & 0\\
    0 & 0 & 0\\
    1 & 0 & 0
    \end{pmatrix},\ S_+ = \begin{pmatrix}
    0 & 0 & 0\\
    0 & 0 & 1\\
    0 & 0 & 0
    \end{pmatrix},\nonumber
\end{equation}
for the fermionic ones. Here we shall consider the decomposition $\lieg = \liem \oplus \liep$
with the Lie superalgebra $\liem$ spanned by $P, Q_+$ and $Q_-$. The corresponding superspace
$\mathcal{M}$ is generated by one bosonic variable $u$ along with two Grassmann variables
$\theta$ and $\bar \theta$. In this case the supergroup element $m$ we introduced above takes
the following matrix form
\begin{equation} \label{eq:m-1d}
m(x) = e^{u P + \theta Q_+ + \bar \theta Q_-} = \begin{pmatrix}
                                                  1 & X & \theta \\
                                                  0 & 1 & 0 \\
                                                  0 & -\bar\theta & 1
                                                \end{pmatrix} \ ,
\end{equation}
where $X = u-\frac12 \theta \bar \theta$ and $x = (u,\theta,\bar\theta)$ represents the three
generators of the structure algebra.
\smallskip

The construction we outlined above provides us with an action of the superconformal algebra
$\lieg$ on this superspace with differential operators $\mathcal{R}_X = u$ of the form
\begin{align}
    & p = \partial_u\ ,\quad & k = -u^2\partial_u - u\theta\partial_{\theta} -
    u\bar\theta\partial_{\bar\theta}\ , \label{eq:sldop1} \\[2mm]
    & d = u\partial_u + \frac12\theta\partial_{\theta} + \frac12\bar\theta
    \partial_{\bar\theta}\ ,\quad
    & r =\theta\partial_{\theta} - \bar\theta\partial_{\bar\theta}\ ,
     \label{eq:sldop2}\\[2mm]
    & q_+ = \partial_{\theta} - \frac12\bar\theta\partial_u \ ,\quad &
    q_- = \partial_{\bar\theta} - \frac12\theta\partial_u\ ,
    \label{eq:sldop3} \\[2mm]
    & s_+ = -(u+\frac12\theta\bar\theta)q_+\ ,\quad &
    s_- = (u-\frac12\theta\bar\theta)q_-\ .
    \label{eq:sldop4}
\end{align}
As we pointed out in our discussion above, the choice of $p$ is not relevant for
the final result. We encourage the reader to derive these explicit expressions
from our general formula \eqref{eq:resultRM}.

\subsection{Global superconformal symmetry and Weyl inversions}

Having constructed superspace along with an action of the superconformal algebra
thereon, our next task is to construct the action of global conformal transformations.
As we shall see in a moment, most of the global transformations act in an obvious
way. The only exception are special conformal transformations. For bosonic conformal
symmetry, the easiest way to construct these is through the conformal inversion of
translations. We follow essentially the same strategy in the supersymmetric context,
except we need to replace the conformal inversion by a closely related Weyl inversion.
The latter extends nicely to superconformal algebras while conformal inversions may
not actually exist, see below.

Defining the action of global conformal transformations on superspace requires a
little bit of preparation. We shall think of a global symmetry transformation as
being associated to a supergroup element $h=h(s)$. We may consider $h$ as a matrix
whose matrix elements are functions on the supergroup, i.e.\ elements of the structure
algebra generated by the coordinates $s_a$ and $s_\alpha$. The graded commutative
algebra that is generated by these coordinates is just another copy of the algebra
that is generated by $x_a$ and $x_\alpha$. From now on we shall suppress the
dependence on $s$ again. The left action of such an element $h$ on the supergroup
element $g(x)= m(x_a)p(x_\alpha)$ is simply given by the left multiplication $g(x)
\mapsto h g(x)$. In order to obtain the action on superspace, we need to
factorize $h g(x)$ as
\begin{equation}
h g(x) = m(y(x,h)) p(x,h) = e^{y(x,h)_a X^a} p(x,h) \ .
\end{equation}
This factorization defines the $h$ transform $h(x)_a$ of the superspace
coordinates $x_a$. Note that $y(x,h)_a$ are elements in the tensor product of
two structure algebras, the one generated by coordinates $x$ and the one that
is generated by $s$. It is particularly easy to apply this definition to
rotations, dilations and R-symmetries since these form a subgroup $K$
that respects the split of $\lieg$ into $\liem$ and $\liep$. In fact, the
Lie algebra $\liek$ is even a subalgebra of $\liep$. In order to factorize
$$k g(x) = k m(x) p(x) = m(y(x,k)) p(x,k)$$
for some $k \in K$\footnote{Here we assume that all the matrix elements are
constant functions on the supergroup, i.e. they are proportional to the
identity element of the structure algebra.} all we need to do is move
$k$ through $m$. Since the generators $X^a$ transform in some representation
$\kappa$ of $K$, the effect can be captured by a linear transformation of the
coordinates $x_a$, i.e. $y(x,k)_a =\kappa_{ab}(k) x_b$. Also (super-)translations
are easy to discuss. These are associated with elements $h(c) = m(c)$ so that
multiplication of $h$ with $g(x)$ only requires to multiply $m(c) m(x) =
m(y(x,c))$. Since bosonic translations commute among each other and with the
supercharges $Q$, the only non-trivial terms in the product $m(c) m(x)$ come
from the non-vanishing anti-commutators of the supercharges. But these can be
evaluated easily in concrete examples and hence the computation of $c(x)$ is
straightforward.
\medskip

It  now remains to discuss the action of special (super-)conformal transformations.
We will not discuss these directly but instead focus on one particular global
superconformal transformation, namely the superconformal extension of the Weyl
inversion $w$. As we shall see, this Weyl inversion relates special super
conformal transformations to supertranslations, just as in the bosonic case.

Before we enter the discussion of the Weyl inversion, let us briefly recall
how the ordinary inversion of conformal field theories is constructed. By
definition, the \textit{conformal group} is a Lie group with $\mathfrak{g}=
\mathfrak{so}(d+1,1)$ as its Lie algebra. Let $O(d+1,1)$
be the group of pseudo-orthogonal matrices. Its identity component is
denoted by $SO^+(d+1,1)$. This group can be realised as the quotient
\begin{equation}
    SO^+(d+1,1) = \Spin(d+1,1)/\mathbb{Z}_2 \nonumber
\end{equation}
of the universal covering group $\Spin(d+1,1)$ by its centre. Both $SO^+(d+1,1)$ and
$Spin(d+1,1)$ act on the compactified Euclidean space, but only the first action is
faithful. In the case of $\Spin(d+1,1)$, both elements of the centre act trivially.
Obviously, both $SO^+(d+1,1)$ and $\Spin(d+1,1)$ possess the same Lie algebra $\lieg
= \mathfrak{so}(d+1,1)$. The conformal inversion
\begin{equation}
    I x^\mu = \frac{x^\mu}{x^2} \nonumber
\end{equation}
is an element of $O(d+1,1)$, but it resides in a component that it not connected to the
identity component, i.e. the conformal inversion $I$ is not an element of $SO^+(d+1,1)$.
We can improve on this issue by multiplying the inversion with some spatial reflection.
The so-called Weyl inversion $w=s_{e_d}\circ I$ involves the reflection on $\mathbb{R}^d$
that sends $x_d$ to $- x_d$ and it belongs to $SO^+(d+1,1)$. We can actually construct
the Weyl inversion explicitly through the following exponential of conformal generators,
\begin{equation}
    w = e^{\pi\frac{K_d-P_d}{2}}. \label{Weyl-inversion}
\end{equation}
There are two elements of $\Spin(d+1,1)$ which project to $w$. We  use the
expression \eqref{Weyl-inversion} as our definition of the Weyl inversion for
$\Spin(d+1,1)$. One can check that its square is the non-trivial element of
the centre, i.e.\ that $w^2=-1$.
\medskip

In passing to the superconformal algebra we use the same formula \eqref{Weyl-inversion} to
define the Weyl element and hence the Weyl inversion. The bosonic part $\lieg_{\bO}$ of the
superconformal algebra $\lieg$ is generated by the bosonic conformal algebra
$\lieg_\textit{bos}$ along with the generators $U \in \mathfrak{u}$ of R-symmetry
transformations. The latter commute with all elements of $\lieg_{\bO}$ and hence the
associated universal enveloping algebras satisfy $U(\lieg_{\bO}) \cong U(\lieg_\textit{bos})
\otimes U(\mathfrak{u})$. By construction $w$ lies in $w \in U(\lieg_\textit{bos})$ and it is
trivial in the $U(\mathfrak{u})$,
$$ w = w \otimes e \in U(\lieg_\textit{bos}) \otimes U(\mathfrak{u}) \cong
   U(\lieg_{\bO})\ . $$
While the action of the element $w$ on generators of the R-symmetry transformations
is trivial, its action on the fermionic generators is not. Using that conjugation of
the generator $D$ of dilations with the Weyl inversion is given by $\text{Ad}_{w_{bos}}
(D)=-D$ we obtain
\begin{equation}
    \frac12\text{Ad}_w(Q) = \text{Ad}_w([D,Q]) = [\text{Ad}_w(D),\text{Ad}_w(Q)] =
    - [D,\text{Ad}_w(Q)]\ , \nonumber
\end{equation}
i.e.\ when a supercharge $Q$ is acted upon by the Weyl inversion it is sent to a generator
whose conformal weight is $-1/2$. Consequently, the Weyl inversion interchanges generators
of supertranslations and super special conformal transformations. For superconformal
algebras of type I, see the final paragraph of the previous subsection for a definition,
one can similarly use that $\text{Ad}_w(R) = w R w^{-1} = R$ to deduce
\begin{equation}
    \text{Ad}_w(\mathfrak{q}_\pm) \subset \mathfrak{s}_\pm \ . \label{odd-generators}
\end{equation}
In conclusion we have seen that the super Weyl inversion exists for all superconformal
algebras and we stated some of its most important properties. This is to be contrasted
with the fact that a supersymmetric analogue of the ordinary conformal inversion may
actually not exist. Assuming that one could choose the superconformal group such that
the inversion $I$ belonged to the bosonic conformal subgroup, then the arguments
leading to eq.\ $(\ref{odd-generators})$ with $w\times e$ replaced by $I\times e$ would
remain valid. On the other hand, as the example $\mathfrak{g} = \mathfrak{sl}(4|1)$ shows,
the fact that $I$ commutes with rotations is inconsistent with eq. $(\ref{odd-generators})$,
bearing in mind that $\mathfrak{q}_+$ and $\mathfrak{s}_+$ are non-isomorphic modules of the
rotation group. Fortunately for us, the existence of the super Weyl inversion will
suffice.

\medskip

\noindent
{\bf Example:} Let us briefly discuss super-conformal transformations and in
particular the super Weyl inversion for the Lie superalgebra $\mathfrak{sl}(2|1)$.
As we discussed at the end of the previous subsection, this Lie superalgebra admits
a 3-dimensional representation. All generators have been spelled out in this
representations above. Within this representation, the supergroup element
$m(x)$ takes the form \eqref{eq:m-1d}. The subgroup $K$ is generated by dilations
and $U(1)_R$ symmetry transformations which are generated by $D$ and $R$, i.e.\ $k
= \exp(\lambda D + \vartheta R)$. Under global transformations with elements $k \in K$
the superspace coordinates $x=(u,\theta,\bar \theta)$ transform as
\begin{equation}
y(x,k) = (e^\lambda u, e^{\frac12 \lambda +\vartheta}\theta, e^{\frac12\lambda - \vartheta} \bar \theta) \ .
\end{equation}
Here we can either think of $\lambda$ and $\vartheta$ as some real parameters of the
transformation or as coordinates on the supergroup, i.e.\ as two generators of the
structure algebra. Supertranslations with an element $m(c) = m(v,\eta,\bar \eta)$
act as $m(c) m(x) = m(c(x))$ with
\begin{equation}
y(x,c) = c(x) = (u+v + \frac12 \theta \bar \eta + \frac12 \bar \theta \eta, \theta+\eta,
\bar \theta + \bar \eta)\ .
\end{equation}
The components of $c = (v,\eta,\bar \eta)$ are generators of the structure algebra.
It remains to discuss the Weyl inversion. Within the 3-dimensional representation
it is straightforward to compute the Weyl inversion from eq.\ \eqref{Weyl-inversion},
\begin{equation} \label{eq:wmatrix}
    w = e^{\pi\frac{K-P}{2}} = \begin{pmatrix}
    0 & -1 & 0\\
    1 & 0 & 0\\
    0 & 0 & 1
    \end{pmatrix}.\nonumber
\end{equation}
Note that $w^2 = \textit{diag}(-1,-1,1)$, i.e.\ it squares to $-1$ within the
bosonic conformal group and is trivially extended within the R-symmetry group.
It is now straightforward to compute the action of the Weyl inversion on superspace
by decomposing the matrix $w m(x) = m(w(x)) p(x,w)$ with $w(x)
= y(x,w)$ given by
\begin{equation}
   \ w(u) = -\frac{1}{u}\ ,\quad  w(\theta) =  \frac{\theta}{u}\ ,\quad
   w(\bar\theta) = \frac{\bar\theta}{u}\ . \label{w-action-1d}
\end{equation}
Note that the action of $w$ on the bosonic coordinate $u$ is the same as in
bosonic conformal field theory. This had to be the case, since in the chosen coordinate
system on the superspace $\mathcal{M}$ the action of the conformal algebra generators on
$x$ is the same as in bosonic theory. Furthermore, we have $w(p,q_+,q_-)w^{-1}=(-k,-s_+,s_-)$,
in accordance with the relations $w^{-1}(P,Q_+,Q_-)w = (-K,-S_+,S_-)$ satisfied by $3\times3$
matrices. Often such conditions are used to derive the action of the inversion. In the
approach here, this is not necessary as the action of $w$ can be computed directly.
This concludes our discussion of the 1-dimensional $\mathcal{N} =2$ superspace and
the global action of the superconformal symmetry on it.

\section{Lifting Correlators to the Supergroup}

This section contains the first new result of the present work. We  establish an isomorphism
between the solutions of superconformal Ward identities that are satisfied by a four-point
function of arbitrary spinning fields and certain covariant functions on the superconformal
group, to which one may also refer as $K$-spherical functions. The construction finds
roots in ideas from \cite{Dobrev:1977qv}, and generalises that of \cite{Buric:2019dfk} to
the superconformal setting. One key ingredient in our formula is a family of supergroup
elements $g(x_i)$ that depends on the insertion points of the four fields in superspace.
These will play an important role in the following sections as well. In the first subsection
we state all this precisely before we illustrate the formulas at the example of $\lieg=
\mathfrak{sl}(2|1)$ in the second. The third subsection contains the proof of our
statement.

\subsection{Statement of the result}

Let us now consider a four-point function in some superconformal field theory.
To each field we associate a copy of our superspace $\mathcal{M}$. The generators
$x_{ia}$ of these spaces carry a label $i=1, \dots, 4$ in addition to
the label $a$ we introduced in the previous section. The corresponding supergroup
elements $m_i = m(x_i)$ are given by
\begin{equation}
m(x_i) = e^{x_{ia} X^a}  .
\end{equation}
Here the summation over $a$ is understood. Given any pair of labels $i,j$ we
define the variables $x_{ij} = (x_{ija}) \in \mathcal{M}_i \otimes \mathcal{M}_j$
through
\begin{equation} \label{eq:xij}
m(x_{ij}) = m(x_j)^{-1} m(x_i) \ .
\end{equation}
Concrete expressions for the components of $x_{ij}$ can be worked out from the
anti-commutator relations of the supercharges $Q$. One may think of $m(x_i)$ as
a function on superspace with values in the universal enveloping algebra or,
more concretely, after evaluation in a fundamental representation of the Lie
superalgebra $\lieg$, as a matrix valued function on superspace.

In the last section we also introduced the Weyl element $w$ through equation
\eqref{Weyl-inversion}. Note that $w$ is constructed out of generators of the
bosonic conformal group only. In particular it acts trivially within the
R-symmetry group $U$. We can think of $w$ as a grouplike element in the
universal enveloping algebra or, after application of a fundamental
representation, as a concrete matrix such as in eq.\ \eqref{eq:wmatrix}.
With the help of the Weyl inversion, let us define a new family of
supergroup elements $n$ through
\begin{equation} \label{eq:nx}
n(x) = w^{-1} m(x) w\ .
\end{equation}
Since $m$ involves only generators $X^a \in \lieg_{>0}$ of the superconformal
algebra that raise the conformal weight, i.e.\ generators $P$ of translations
and supercharges $Q$, the element $n$ is built using generators $Y^a$ from the
algebra $\lieg_{<0}$ that lower the conformal weight, see our previous discussion
of the Weyl inversion. This means that $n$ involves special conformal generators
$K$ as well as the fermionic generators $S$.

In order to proceed, let us introduce another supergroup element $k=k(t)$ using
the remaining generators $X \in \lieg_0$ that commute with the generator of
dilations and therefore neither appear in $n$ nor in $m$. It means that $k$ is
built from the generators of dilations, rotations and R-symmetry
transformations, all of which are even (bosonic). Given the three supergroup
elements $m,n,k$ we can now decompose $w m(x)$  as
\begin{equation} \label{eq:factorization}
w m(x) = m(y(x))\,  n(z(x)) \, k(t(x)) \ ,
\end{equation}
where the components of $y(x) = (y(x)_a)$, $z(x) = (z(x)_a)$ and $t(x)
= (t(x)_\varrho)$ are certain functions of the superspace coordinates $x_i$ that
can be worked out concretely on a case-by-case basis. We shall state concrete
formulas in some examples below. Let us stress that it is through this factorization
\eqref{eq:factorization} that we introduce the action of the Weyl inversion
$w$ on superspace, i.e. by definition $y(x) = w x$. We  consider the
functions $y,z$ and $t$ as given for now and use them to introduce
\begin{equation}
y_{ij} = y(x_{ij}) = w x_{ij}  \ , \quad  z_{ij}= z(x_{ij}) \ , \quad
t_{ij} = t(x_{ij})  \ .
\end{equation}
By definition we have
\begin{equation} \label{eq:factorizationij}
w m(x_{ij}) = m(y_{ij}) \, n(z_{ij}) \, k(t_{ij}) \ .
\end{equation}
The components of $x_{ij}, y_{ij}, z_{ij}$ and $t_{ij}$ are elements in the
four-fold tensor product $\mathcal{M}^4 \cong \mathcal{M}^{\otimes_4}$ of the
superspace $\mathcal{M}$, one copy for each insertion point. This is all we
need to know about the superconfiguration space of the four insertion points.
\medskip

So, let us now consider some four-point correlation function $G$ in a quantum
field theory with superconformal symmetry given by $\lieg$. The fields $\Phi$ of our
theory are organized in supermultiplets. We label these supermultiplets through
the quantum numbers of their superprimaries. These consist of a conformal weight
$\Delta$, a spin $\lambda$ and the R-charges $q$. The collection of these
quantum  numbers determine a finite dimensional irreducible representation
$\rho = \rho_{\Delta, \lambda,q}$ on the Lie algebra $\mathfrak{k} = \lieg_0$
that is spanned by dilations, rotations and R-symmetries. We denote the carrier
space of this representation by $V = V_\rho$ and shall often refer to it as the
space of superpolarizations. Let us stress that elements of $V_\rho$ are associated
with polarizations of the superprimary in the supermultiplet $\Phi$. In our
four-point function we have four supermultiplets whose superprimary components
transform in representations $\rho_i,\ i = 1, \dots, 4$. The polarizations of
these four superprimary fields span the vector spaces $V_i$.

Given these data, we now consider the space $\mathcal{F}(\mathfrak{g})\otimes
V_{1234}$ of ``functions $F$ on the supergroup'' that take values in the vector
space $V_{1234} = V_1 \otimes \dots \otimes V_4$. Among its elements we restrict
to those functions $F$ that possess the following covariance property\footnote{
Mathematically minded readers should think of $g$ as a supergroup element $g \in
U(\lieg) \otimes \mathcal{F}$ where $\mathcal{F}$ can be any graded commutative
algebra, see section 2.1. The object $F(g) \in \mathcal{F} \otimes V_{1234}$ is
then obtained using the duality between $\mathcal{F}(\lieg)$ and $U(\lieg)$,
see eq.\ \eqref{eq:duality}.}
\begin{align}\label{eq:covariance}
    F(k_l g k_r)= \Big(\rho_1(k_l)\otimes\rho_2(w k_l w^{-1})\otimes
    \rho_3(k_r^{-1})\otimes\rho_4(w k_r^{-1}w^{-1})\Big) F(g) \ ,
\end{align}
for all $k_l,k_r \in K$. In analogy with ordinary Lie theory, such an $F$ will
be called a $K$-spherical function. To digest the mathematical meaning of this
formula a bit better, let us pretend for a moment that we are dealing with some
ordinary Lie algebra $\lieg$ rather than a superalgebra. In that case, $g$ as
well as $k_l, k_r$ are elements of the bosonic group $G$. When we write $F(g)$
we let the group element $g$ act as a global symmetry transformation on the
space $\mathcal{F}(\lieg)$ of functions on the group and evaluate the result
at the group unit. Stated more directly we simply evaluate the vector valued
function $F$ at the point $g$ of the group manifold. Almost the same is true
for superalgebras except that $g$ is a matrix whose matrix elements are taken
from some Grassmann algebra and $F$ is a prescription that turns such a matrix
into a vector $F(g)$ whose components are elements of that Grassmann algebra.
To evaluate $F(k_l g k_r)$ we employ the left-right action of $K \times K$ on
the space $\mathcal{F}(\lieg)$ of functions on the supergroup to transform $F$
into new element of $F^{(k_l,k_r)}$ of the space $\mathcal{F}(\lieg) \otimes
V_{1234}$. When we apply this transformed $F^{(k_l,k_r)}$ to $g$ we obtain
another vector $F^{(k_l,k_r)}(g) = F(k_lgk_r)$ with Grassmann valued components.
The covariance condition \eqref{eq:covariance} selects those elements $F$ for
which the two vectors $F(g)$ and $F(k_lg k_r)$ are related by a specific matrix
rotation that is obtained from representation matrices of $k_l$ and $k_r$ in the
representations $\rho_i$. The precise construction of this matrix, which
also involves conjugation with the Weyl element $w$ in two of the four
tensor factors will become clear in the third subsection.

Let us now come back to our correlation function $G_4$. By construction, $G_4(x_i)$
is a function on the four-fold tensor product $\mathcal{M}^4$ of superspace that takes
values in the space $V_{1234}$ of polarizations, i.e.\ $G_4 \in \mathcal{M}^4 \otimes
V_{1234}$. Being the four-point function in some superconformal field theory, $G_4$
transforms in a very special way under superconformal transformations. This can be
expressed in terms of a set of superconformal Ward identities. As a consequence of
these covariance properties one may show that, given $G_4$, there exists a unique
function $F \in \mathcal{F}(\lieg) \otimes V_{1234}$ on the supergroup with
covariance property \eqref{eq:covariance} such that
\begin{eqnarray}
G_4(x_i) & = & \Big(1\otimes\rho_2(k(t_{21}))^{-1}\otimes1\otimes
\rho_4(k(t_{43}))^{-1}\Big) F(g(x_i))\, ,  \label{magic-formula}\\[2mm]
& & \textit{where}\ g(x_i) =  n(y_{21})^{-1} m(x_{31}) n(y_{43})\ .  \label{eq:gxi}
\end{eqnarray}
The argument of $F$ is a product of supergroup elements, i.e.\ an element of $U(\lieg)
\otimes \mathcal{M}^4$ or some matrix representation thereof. After the application of
$F$ we obtain an element of $\mathcal{M}^4 \otimes V_{1234}$. We may think of this as
a vector valued function on the four-fold tensor product of superspaces which can be
compared to $G_4$. The factor in front of $F$, that relates $F(g(x_i))$ to $G_4(x_i)$
is a certain matrix of functions on $\mathcal{M}$ that acts non-trivially on the two
factors $V_2$ and $V_4$. We shall also refer to eq.\ \eqref{magic-formula} as the
supersymmetric \textit{lifting formula}.
\smallskip

Let us remark that there is a quick sanity check of our formula, namely one may
verify that both sides of the lifting formula \eqref{magic-formula} satisfy the same
Ward identities for infinitesimal transformations generated by elements in $X
\in \lieg_{\geq 0}$. The latter is spanned by translations, supercharges $Q$,
rotations, dilations and R-symmetry  transformations. The key observation is
that
\begin{equation}\label{eq:diffrel}
\sum_{j=1}^4 \mathcal{R}_X^{(j)} g(x_i)
= \left[ X \otimes \id, g(x_i) \right]\ .
\end{equation}
Recall that the argument $g(x_i)$ of $F$ may be considered as a matrix whose
entries are functions on the four-fold product of superspace. On these matrix
elements we act with the sum of right invariant vector fields $\mathcal{R}_X$
for $X \in \lieg_{\geq 0}$, acting on one set of superspace coordinates each.
The differential operators $\mathcal{R}$ were constructed in the previous section.
Our claim is that the resulting matrix of functions on superspace is the same
as for the matrix commutator of the representation matrix for $X$ with the
product of supergroup elements. This property holds essentially by construction
of the argument of $F$. This is not a full proof of our formula yet since the
argument cannot easily be extended to special (super-)conformal transformations.
We  give a complete derivation in the third subsection after we have
illustrated the notations and constructions we introduced in this section
for the $\mathcal{N}=2$ superconformal algebra in $d=1$ dimension.

\subsection{Illustration for 1-dimensional superconformal algebra}

Let us continue to illustrate our constructions and statements in the example of the
$\mathcal{N}=2$ superconformal algebra in $d=1$. Recall that the fundamental representation
of this algebra is 3-dimensional and hence we realize all our supergroup elements as
$3\times 3$ matrices with components in the superspace. The elements $m(x)$ were
constructed in eq.\ \eqref{eq:m-1d} already. The Weyl inversion $w$ and its action on
superspace were worked out in eqs.\ (\ref{eq:wmatrix}) and \eqref{w-action-1d},
respectively. It is easy to determine the $3 \times 3$ matrices $n(x)$ to take
the form
\begin{equation} \label{eq:n-1d}
    n(x) = w^{-1} m(x) w = \begin{pmatrix}
    1 & 0 & 0\\
    -X & 1 & -\theta\\
    -\bar\theta & 0 & 1
    \end{pmatrix}\ ,
\end{equation}
where $X = u - \frac12 \theta \bar\theta$ is the same even combination of
superspace coordinates $(u,\theta,\bar \theta)$ that appeared in our formula
\eqref{eq:m-1d} for $m(x)$. The central ingredient in our construction above is
the factorization formula \eqref{eq:factorization} for $w m(x)$. In the case
of $\lieg = \mathfrak{sl}(2|1)$ this reads
\begin{equation}
    \begin{pmatrix}
    0 & -1 & 0\\
    1 & X & \theta\\
    0 & -\bar\theta & 1
    \end{pmatrix} = \begin{pmatrix}
    1 & -\frac1u \left(1+\frac{\theta\bar\theta}{2u}\right) & \theta/u\\
    0 & 1 & 0\\
    0 & -\bar\theta/u & 1
    \end{pmatrix} \begin{pmatrix}
    1 & 0 & 0\\
    u+\frac12\theta\bar\theta & 1 & \theta\\
    \bar\theta & 0 & 1
    \end{pmatrix} \begin{pmatrix}
    \frac1u\left(1-\frac{\theta\bar\theta}{2u}\right) & 0 & 0\\
    0 & u\left(1-\frac{\theta\bar\theta}{2u}\right) & 0\\
    0 & 0 & 1-\frac{\theta\bar\theta}{u}
    \end{pmatrix} \ . \label{eq:matrixfactorization-1d}
\end{equation}
Comparing the first of the three factors with the expression \eqref{eq:m-1d} for
$m(y)$ we deduce
\begin{equation}
y(x) = (Y+\frac12\eta\bar\eta,\eta,\bar\eta) = w(u,\theta,\bar\theta) =
\left(\frac{-1}{u},\frac{\theta}{u},\frac{\bar\theta}{u}\right)\ . \label{eq:wact-1d}
\end{equation}
This agrees of course with the result we found in eq. \eqref{w-action-1d}. Turning
to the second matrix factor in the factorization formula and comparing with eq.\
\eqref{eq:n-1d} for $n(z)$ we conclude
\begin{equation} \label{eq:zcoord-1d}
z(x) = (Z+\frac12\zeta\bar\zeta,\zeta,\bar\zeta) = (-u,-\theta,-\bar\theta) \ .
\end{equation}
Using the representation matrices for $D$ and $R$ that we spelled out in
eq.\ \eqref{eq:bosrep}, the third factor, finally, can be written as
\begin{equation}
k(t(x)) = e^{-\log u^2 D + \frac{\theta\bar\theta}{2u}R}\ .
\end{equation}
We lift the matrix equation \eqref{eq:matrixfactorization-1d} to the following
factorization identity for supergroup elements
\begin{equation}
  w m(x) =  w e^{x\cdot X} =
  e^{w(x)\cdot X} e^{-x\cdot X^w} e^{-\log u^2 D + \frac{\theta\bar\theta}{2u}R} \ ,
  \label{fund-1d}
\end{equation}
where $X^w = w^{-1}(P,Q_+,Q_-)w = (-K,-S_+,S_-)$. Given several points $x_i$ in superspace,
we can now compute the supercoordinates $x_{ij}$ by evaluating the product $m(x_j)^{-1}
m(x_i)$. The result is given by $x_{ij} = (u_{ij},\theta_{ij}, \bar \theta_{ij})$ with
\begin{equation}
    u_{ij} = u_i - u_j -\frac12\theta_i\bar\theta_j - \frac12\bar\theta_i\theta_j \ ,\quad
    \theta_{ij} = \theta_i - \theta_j \ ,\quad
    \bar\theta_{ij} = \bar\theta_i - \bar\theta_j \ .
    \label{distance}
\end{equation}
For completeness let us also state how the Weyl inversion acts on $x_{ij}$
\begin{equation}
    w(x_{ij}) = (-u_{ij}^{-1},u_{ij}^{-1}\theta_{ij},
        u_{ij}^{-1}\bar\theta_{ij})\ . \label{inverse}
\end{equation}
Of course this coincides with the formula \eqref{eq:wact-1d} applied to the
superspace coordinates $x_{ij}$. At this point we have explained all the
ingredients that are needed to construct the supergroup elements $g(x_i)$
that were introduced in eq.\ \eqref{eq:gxi}.
\smallskip

Let us now consider a four-point function $G_4$ of primary fields with conformal
weights $\Delta_i$ and R-charges $r_i$ for $i=1, \dots, 4$. Given $\Delta$ and
$r$, the corresponding representation $\rho_i$ of the group $K = SO(1,1) \times
U(1)$ reads
\begin{equation} \label{eq:rho-1d}
    \rho_{\Delta,r}(e^{\lambda D + \kappa R}) = e^{-\Delta\lambda + r\kappa}\ .
\end{equation}
Since the group $K$ is abelian, the space $V$ of polarizations is 1-dimensional
and so is the tensor product $V_{1234} = V_1 \otimes \dots \otimes V_4$. According
to our general result \eqref{magic-formula}, there exists a unique functional $F$
with the covariance properties
\begin{align}
    F(e^{\lambda_l D + \kappa_l R} g e^{\lambda_r D + \kappa_r R})=
    e^{(\Delta_2-\Delta_1)\lambda_l+(r_1+r_2)\kappa_l} e^{(\Delta_3-\Delta_4)\lambda_r
    - (r_3+r_4)\kappa_r} F(g) \ ,
\end{align}
such that the lifting formula reads
\begin{equation}
    G_4(x_i) = \Omega(x_i) \, F(e^{-w(x_{21})\cdot X^w} e^{x_{31}\cdot X}
    e^{w(x_{43})\cdot X^w})
    \label{magic-1d}
\end{equation}
and the prefactor $\Omega$ is given by
\begin{equation} \label{eq:Omegasl2}
\Omega =\Omega(x_i) = \frac{e^{r_2\frac{\theta_{12}\bar\theta_{12}}{2u_{12}}+
    r_4\frac{\theta_{34}\bar\theta_{34}}{2u_{34}}}}{u_{12}^{2\Delta_2}
    u_{34}^{2\Delta_4}}\ .
\end{equation}
It is instructive to verify that the commutation relations \eqref{eq:diffrel}
hold for the argument of $F$ and to evaluate $G_4$ for Weyl-inverted arguments
$w(x_i)$, thereby showing that the right hand side of the lifting formula
\eqref{magic-1d} indeed satisfy the same conformal Ward identities as the
four-point function.

\subsection{Proof of the lifting formula}

The goal of this subsection is to prove the main result \eqref{magic-formula} for
an arbitrary superconformal group. Before doing that, let us give one more definition,
an extension of the factorization formula \eqref{eq:factorization}
\begin{equation}
    h m(x) = m(y(x,h)) n(z(x,h)) k(t(x,h)) \ , \label{matrix-identity}
\end{equation}
from the Weyl inversion $h = w$ to arbitrary elements $h$ of the superconformal
group. This formula also extends our analysis in section 2.3 where we studied the action
of global conformal transformations on superspace. At the time we only cared about the
first factor $m(y(x,h))$ in the product on the right hand side. The new formula
\eqref{matrix-identity} extends the action of global superconformal transformations
to the whole superconformal group. Otherwise all the additional explanations we
provided in section 2.3. remain applicable. The extended factorization formula involves
three sets of functions $y(x,h) = (y(x,h)_a)$, $z(x,h) = (z(x,h)_a)$ and $t(x,h) =
(t(x,h)_\varrho)$. For $h=w$ we recover the functions we introduced in the
previous section.

A four-point correlation function $G_4$ satisfies a set of Ward identities. For global
superconformal transformations $h$ these may be written in the form
\begin{equation} \label{eq:G4Wardid}
    G_4 (x_i^h) = \Big(\bigotimes_{i=1}^4 \rho_i (k(t(x_i,h)))\Big) G_4(x_i) \ .
\end{equation}
Note that correlation functions are essentially invariant under these transformations
except some factors depending in the weight, spin and the R-charges. This dependence
is encoded in the choice of representations $\rho_i$, as we explained above. In a first
step we want to lift the correlator $G_4$ to and object $F_4\in\mathcal{F}_1\otimes V_1
\otimes \dots \otimes \mathcal{F}_4\otimes V_4$, where $\mathcal{F}_i$ are supercommuting
copies of the structure algebra $\mathcal{F}(\mathfrak{g})$ of functions on the supergroup.
This can be done in a unique way if we require
\begin{equation}\label{eq:F4rightcov}
    F_4(m(x_i)) = G_4(x_i)\  ,\quad \quad F_4 (g_i n_i k_i) =
    \bigotimes_{i=1}^{4} \rho_i (k_i^{-1})
    F_4(g_i) \ .
\end{equation}
Here our notations are the same as in section 3.1, see our extended discussion before
equation \eqref{magic-formula}. The Ward identities \eqref{eq:G4Wardid} satisfied by
$G_4$ imply the following invariance conditions satisfied by $F_4$ under simultaneous
left multiplication of its four arguments by an element $h$ of the superconformal group,
\begin{eqnarray}\label{eq:F4leftinv}
    F_4(h m(x_i)) & = & F_4\Big( m(x_i^h) n(z(x_i,h)) k(t(x_i,h)) \Big) \\[2mm]
    & = & \Big( \bigotimes_{i=1}^4 \rho_i (k(t(x_i,h))^{-1}) \Big) G_4(x_i^h) =
    G_4(x_i) = F(m(x_i)) \ .
\end{eqnarray}
Other than the Ward identity, we have used the definitions $(\ref{matrix-identity})$
and $(\ref{eq:F4rightcov})$. Given this element $F_4$ and the Weyl inversion $w$ we
can construct a new object $F\in\mathcal{F}(\mathfrak{g})\otimes V_{1234}$ through
the prescription
\begin{equation}\label{eq:FfromF4}
    F(g) := F_4 (e,w^{-1},g,gw^{-1}) \ .
\end{equation}
While this might look a bit bizarre at first, it is easy to verify that it
defines a $K$-spherical function $F$, i.e. that $F$ satisfies the covariance
law \eqref{eq:covariance}. Indeed, from the definition \eqref{eq:FfromF4} of
$F$, the left invariance condition \eqref{eq:F4leftinv} and the right
covariance law in eq.\ \eqref{eq:F4rightcov} of $F_4$ we obtain
\begin{align*}
    F(k_l g k_r) &= F_4(e,w^{-1},k_l g k_r, k_l g k_r w^{-1}) =
    F_4(k_l^{-1},w^{-1} w k_l^{-1}w^{-1},g k_r, g w^{-1} w k_r w^{-1} )\\[2mm]
                 &= \Big(\rho_1(k_l)\otimes\rho_2(w k_l w^{-1})
                 \otimes\rho_3(k_r^{-1})\otimes\rho_4(wk_r^{-1}w^{-1})\Big) F(g) \ .
\end{align*}
In conclusion we have shown that a correlation function $G_4$ provides us with a
$K$-spherical function $F$. It is actually not difficult to invert the map and
recover $G_4$ from $F$. Suppressing the last two arguments and their corresponding
prefactors for simplicity, we have
\begin{eqnarray*}
F_4(m(x_1),m(x_2)) & = & \left(1 \otimes \rho_2(k(t_{21})^{-1})\right)
F_4\left(m(x_1) n(y_{21}), m(x_2) k(t_{21})^{-1} n(z_{21})^{-1} \right) \\[2mm]
& = & \left(1 \otimes \rho_2(k(t_{21})^{-1})\right)
F_4\left(m(x_1) n(y_{21}), m(x_1) m(x_{21}) k(t_{21})^{-1} n(z_{21})^{-1} \right) \\[2mm]
& = & \left(1 \otimes \rho_2(k(t_{21})^{-1})\right)
F_4\left(m(x_1) n(y_{21}), m(x_1) w^{-1} m(y_{21}) \right) \\[2mm]
& = & \left(1 \otimes \rho_2(k(t_{21})^{-1})\right)
F_4\left(m(x_1) n(y_{21}), m(x_1) n(y_{21}) w^{-1} \right).
\end{eqnarray*}
In the first step we used the covariance property \eqref{eq:F4rightcov} of $F_4$ in the
first two arguments to multiply the first argument with $n(y_{21})$ and the second with
$k(t_{21})^{-1} n(z_{21})^{-1}$. Since the latter contains a factor $k$ it needed to
be compensated by a rotation in the second factor of the space of superpolarizations.
Then we inserted the definition of $m(x_{21})$ and used that
$$ m(x_{21}) = w^{-1} m(y_{21}) n(z_{21}) k(t_{21})\ .  $$
This factorization formula is essentially the definition of $y_{21}, z_{21}$ and $t_{21}$.
Finally we moved the Weyl element $w^{-1}$ through $m$ using that $n = w^{-1} m w$. We can now
apply the same steps to the third and fourth argument to obtain
\begin{equation}\label{eq:F4ggwggw}
F_4(m(x_i)) = \left(1 \otimes \rho_2(k(t_{21})^{-1}) \otimes 1 \otimes \rho_4(k(t_{43})^{-1})\right)
F_4\left(g_{12}(x_i),g_{12}(x_i) w^{-1}, g_{34}(x_i), g_{34}(x_i) w^{-1} \right),
\end{equation}
where we introduced the elements
$$ g_{ij}= m(x_i) n(y_{ji})\ .  $$
Finally, we can use the invariance property \eqref{eq:F4leftinv} of
$F$ for $h = g_{12}^{-1}$ to obtain
$$ F_4(m(x_i)) = \left(1 \otimes \rho_2(k(t_{21})^{-1})
  \otimes 1 \otimes \rho_4(k(t_{43})^{-1})\right)
  F_4\left(e, w^{-1}, g(x_i), g(x_i) w^{-1}\right) \ .
$$
Here $g(x_i)$ is the element we introduced in eq.\ \eqref{eq:gxi}. Using our definition of the functional
$F$ in eq.\ \eqref{eq:FfromF4} and the relation between $F_4$ and $G_4$ we have thereby established the
lifting formula \eqref{magic-formula}.
\medskip

From the above derivation, one may deduce the following transformation properties of $g_{ij}$ and $k(t_{ji})$
under superconformal transformations
\begin{eqnarray}
g_{ij}(x^h) & = & h \, g_{ij}(x)\,  k(t(x_i,h))^{-1} \ , \label{eq:gijtrafoh}\\[2mm]
k(t_{ji}^h) & = & k^{w}(t(x_i,h))\, k(t_{ji})\,  k(t(x_j,h))^{-1}\ , \label{eq:ktrafoh}
\end{eqnarray}
where $k^{w} = w k w^{-1}$. Indeed these are necessary for the right hand
side of eq. \eqref{eq:F4ggwggw} to satisfy the same Ward identities as the left hand side. A complete proof
of the two transformation laws can be found in appendix $A$. These two formulas will play a significant
role in the computation of the crossing factor to which we turn next.

\section{Tensor Structures and Crossing Symmetry Equations}

Having lifted the spinning four-point fucntion $G_4$ from superspace to the
superconformal group through eq.\ \eqref{magic-formula} we can now employ
(super)group theoretic constructions to study superconformal correlators.
In the first subsection we employ a supersymmetric version  of the Cartan
or KAK decomposition for superconformal groups of type I to factorize
four-point functions into the product of a tensor factor $\Theta =
\Theta(x_i)$ and a function $\Psi$ that depends on superconformal cross ratios
only.\footnote{As explained in the introduction, our group theoretic factorization
$G_4 = \Theta \Psi$ is reminiscent of the factorization $G_4 = \Omega g$ used in
most of the CFT literature. The difference between the two factorizations can be
quantified through the ratio $\Theta \Omega^{-1}$ which has a non-trivial
dependence on cross ratios. As we shall see below, $\Theta$ and $\Psi$ are
more universal than the factors $\Omega$ and $g$.} This part of our analysis
extends constructions in \cite{Buric:2019dfk} to the superconformal setting.
We can perform the factorization for different channels. The supercrossing
factor, i.e. the ratio of the corresponding tensor factors $\Theta_s$ and
$\Theta_t$ for the $s$- and $t$-channel, is studied at the end of the first
subsection. There we establish its superconformal invariance and compute it
for bosonic conformal symmetries in any dimension $d$.

At this stage, all quantities depend on fermionic variables. In particular,
the function $\Psi$ still depends on some number of nilpotent invariants.
By expanding all quantities in the Grassmann variables we construct the
crossing factor in the second subsection. This is then used to write the
crossing symmetry constraints in the independent coefficients of the
operator product expansions in terms of functions of two bosonic cross
ratios only. As shown in \cite{Buric:2019rms}, the latter may be expanded
into wave functions of some Calogero-Sutherland Hamiltonian. By collecting
all the material we have put together through our discussion of the
example $\lieg = \mathfrak{sl}(2|1)$ we can finally calculate the
crossing factor for $\mathcal{N}=2$ superconformal field theories in
$d=1$ dimension, see third subsection.

\subsection{Cartan coordinates, tensor and crossing factors}

We will now construct the tensor structures, starting from the lifting formula \eqref{magic-formula}.
Note that eq.\  \eqref{magic-formula} treats each of the four insertion points differently
and hence it breaks the permutation symmetry of correlators in a Euclidean theory. Different
permutations $\sigma$ of the four points are associated with different channels. We  refer to
the channel that is associated with the identity permutation $\sigma = \sigma_s = \textit{id}$
as the $s$-channel. Another important case for us is the permutation $\sigma = \sigma_t = (24)$
which we call the $t$-channel. In any case, given the choice of the channel $\sigma$, we can
extend the lifting formula \eqref{magic-formula} to become
\begin{equation} \label{eq:magic-formulasigma}
G_4(x_i) = \rho_{\sigma(2)}(k(t_{\sigma(2)\sigma(1)})^{-1})
\rho_{\sigma(4)}(k(t_{\sigma(4)\sigma(3)})^{-1}) F_\sigma(g(x_{\sigma(i)}))  \ .
\end{equation}
Here, the factor $\rho_{\sigma(i)}$ acts on the $\sigma(i)^\textit{th}$ tensor factor in
the space of superpolarizations and it acts trivially on all other tensor factors.

In order to proceed we adopt a new coordinate system that we  refer to as
Cartan coordinates. So far we have decomposed supergroup elements $g$ into factors
$m$, $n$ and $k$. Now we  consider a different decomposition in which supergroup
elements $g$ are written as
\begin{equation}  \label{eq:sCartan}
g = k_l \eta_l a \eta_r k_r \ ,
\end{equation}
where $k_{l}$ and $k_{r}$ are associated with the subgroup $K$ that is obtained through
exponentiation of rotations, dilations and R-symmetry transformations. Similarly, the
factors $\eta_l$ and $\eta_r$ are associated with fermionic generators. More specifically,
each of these factors contains half of the generators in the odd subspace $\lieg$. In the
following we  consider Lie superalgebra $\lieg$ of type I for which the internal symmetry
group contains a $U(1)$-factor which allows us to decompose the fermionic generators according
to the sign of the $U(1)$ R-charge. It turns out that half of the supercharges $Q$ possess
positive R-charge while the others possess negative R-charge and similarly for the super
special conformal transformations $S$. Let us agree that $\eta_l$ uses generators of
negative charge while $\eta_r$ is build from generators with positive charge. The central
factor $a = a(u_1,u_2)$, finally, depends on two bosonic coordinates $u_1,u_2$ only and
it is assumed to take the form
\begin{equation}
\label{adef}
a(u_1,u_2) =  e^{\frac{u_1+u_2}{4}(P_1+K_1) -i \frac{u_1-u_2}{4}(P_2 - K_2)}\ .
\end{equation}
Let us note that a factorization of supergroup elements $g$ in the form \eqref{eq:sCartan}
is not unique. In fact, given any such factorization we can produce another factorization
of the very same form by the transformation
\begin{equation} \label{eq:gaugeB}
\left(k_l,\eta_l;k_r,\eta_r\right) \rightarrow
\left(k_l b, b^{-1} \eta_l b ; b^{-1} k_r, b^{-1} \eta_r b\right),
\end{equation}
where $b$ are elements associated with the subalgebra $\mathfrak{so}(d-2)\oplus \mathfrak{u}_r
\subset \liek$ and therefore commute with $a = a(u_1,u_2)$. At the same time, the elements
$b^{-1} \eta_{l/r} b$ can still be written as exponentials of fermionic generators with
negative(l) and positive(r) $U(1)$ R-charge, respectively. Hence our gauge transformation
\eqref{eq:gaugeB} respects the Cartan decomposition. Elements $b$ form the stabilizer
group $B = SO(d-2) \times U_r$ of the Cartan decomposition \eqref{eq:sCartan}. For later
use we introduce a projector $P$ by integrating $b$ over the entire stabilizer group $B$,
\begin{equation}\label{eq:Pdef}
P = \frac{1}{\text{Vol}\  B} \int_B d\mu  b \ =
\frac{1}{\text{Vol}\  B} \int_B d\mu(\beta) b(\beta) \ ,
\end{equation}
where $\mu$ is the Haar measure on $B$. For pedagogical reasons we have introduced some
coordinates $\beta$ on $B$ so that the element $b$ could be written explicitly as a function
$b = b (\beta)$ on $B$ with values in $U(\mathfrak{b})$. As we indicated, $P$ can be considered
as an element in the universal enveloping algebra $U(\mathfrak{g})$. More concretely, after
evaluation in some representation we can also think of $P$ is a matrix. By construction, this
matrix has two important properties
\begin{equation} \label{eq:Pprop}
P^2 = P \quad , \quad b P  = P\ .
\end{equation}
We can verify the second equation very easily using the integral representation of $P$ and
the left invariance of the Haar measure. The first property then follows from $bP = b(\beta)
P$ by performing an additional integration over $B$ since $P$ is a constant on $B$.

In our analysis below we will apply the projector $P$ to an a function $f(u_i,\theta,\bar \theta)$
that takes values in the representation space $V_{1234}$ of $K$. The latter may be considered a
carrier space for a representation of $B$ by restriction from $K$ to its subgroup $B$. The
representation of $P$ on such an object $f$ is denoted by $\mathcal{P}$, i.e.
\begin{equation} \label{eq:calP}
\mathcal{P} [f(u_i,\theta,\bar \theta)]= \frac{1}{\text{Vol}\  B} \int_B d\mu  \chi(b)
f(u_i,\theta^b,\bar\theta^b)\ ,
\end{equation}
where $\theta^b$ and $\bar \theta^b$ denotes the action of $b$ on the Grassmann coordinates
$\theta$ and $\bar \theta$ and $\chi(b)$ is a shorthand for the action of $b$ on the finite
dimensional vector space $V_{1234}$ of superpolarizations,
\begin{equation}  \label{eq:chi}
\chi(b)=\rho_1(b)\otimes \rho_2(w b w^{-1})\otimes \rho_3(b)\otimes \rho_4(w b w^{-1})\ .
\end{equation}
In practical computations it is convenient to make some specific choices for the
Cartan factors that remove the gauge freedom \eqref{eq:gaugeB}. Such gauge fixing
conditions are arbitrary and at the end of every calculation one has to check that
the result does not depend on them.
\medskip

Let us now apply the Cartan factorization to the argument $g(x_{\sigma(i)})$ of the
functional $F_\sigma$ in eq.\ \eqref{eq:magic-formulasigma},
\begin{equation}\label{Cartan-factors}
   g(x_{\sigma(i)}) = k_{\sigma,l}(x_i) \eta_{\sigma,l}(x_i)a_\sigma(x_i)
   \eta_{\sigma,r}(x_i) k_{\sigma,r}(x_i) \ .
\end{equation}
The formula \eqref{eq:magic-formulasigma} and covariance properties of $F_\sigma$ give
\begin{align}
G_4(x_i)  & =  \rho_{\sigma(2)}(k(t_{\sigma(2)\sigma(1)})^{-1})
\rho_{\sigma(4)}(k(t_{\sigma(4)\sigma(3)})^{-1}) F_\sigma(g(x_{\sigma(i)})) \nonumber \\[2mm]
& \hspace*{-4pt} = \rho_{\sigma(2)}\left(k(t_{\sigma(2)\sigma(1)})\right)^{-1} \rho_{\sigma(4)}
\left(k(t_{\sigma(4)\sigma(3)})\right)^{-1} F_\sigma(k_{\sigma,l}\eta_{\sigma,l}
a_\sigma \eta_{\sigma,r}k_{\sigma,r}) \\[2mm]
& \hspace*{-4pt} = \rho_{\sigma(1)}(k_{\sigma,l})\rho_{\sigma(2)}\left(k(t_{\sigma(2)\sigma(1)})^{-1}
k_{\sigma,l}^{w}\right)\rho_{\sigma(3)}(k_{\sigma,r}^{-1})\rho_{\sigma(4)}
\left(k(t_{\sigma(4)\sigma(3)})^{-1} (k^{-1}_{\sigma,r})^{w}\right)
F_\sigma(\eta_{\sigma,l}a_\sigma \eta_{\sigma,r}) . \nonumber
\end{align}
For simplicity, we dropped the dependence of Cartan factors on the insertion points, i.e.\
for example $k_{\sigma,l} = k_{\sigma,l}(x_i) = k_l(x_{\sigma(i)})$. We will
discuss the concrete functional dependence of the insertion points a bit later.

Let us spell out the previous formula for the $s$ and $t$-channel. In the
$s$-channel one obtains
\begin{equation} \label{eq:G4schannel}
G_4(x_i)   =  \rho_{1} (k_{s,l}) \rho_{2}(k(t_{21})^{-1}
k^{w}_{s,l}) \rho_{3}(k_{s,r}^{-1})
\rho_{4}(k(t_{43})^{-1} (k^{w}_{s,r})^{-1})
\mathcal{P}_s F_s(\eta_{s,l}a_s \eta_{s,r})\ ,
\end{equation}
while the $t$-channel gives
\begin{equation} \label{eq:G4tchannel}
G_4(x_i) = \rho_{1} (k_{t,l}) \rho_{4}(k(t_{41})^{-1} k^{w}_{t,l}) \rho_{3}(k_{t,r}^{-1})
\rho_{2}(k(t_{23})^{-1} (k^{w}_{t,r})^{-1}) \mathcal{P}_t F_t(\eta_{t,l}a_t \eta_{t,r})\ .
\end{equation}
Here we introduced projector $\mathcal{P}$ that was defined in eq.\ \eqref{eq:calP} explicitly
to stress that $F(\eta_l a\eta_r)$ takes value in the space of $B$-invariants. Roughly
speaking, the two factors in front of $F_{s}$ and $F_t$ are the $s$- and $t$-channel tensor
structures.

The ratio of these $s$- and $t$-channel tensor structures is referred to as supercrossing
factor and we denote it by $\mathcal{M}$. As we can read off the the previous two formulas
the supercrossing factor takes the form
\begin{equation}  \label{eq:crossingmatdef}
\mathcal{M}_{st}(x_i) = \mathcal{P}_t \, \bigotimes_{i=1}^4 \rho_i(\kappa_i) \, \mathcal{P}_s
\ , \end{equation}
where the four elements $\kappa_i$ are given by
 \begin{eqnarray}
\kappa_1   = k_{t,l}^{-1}k_{s,l} \quad & , & \quad
\kappa_{2} = k^{w}_{t,r} k(t_{23})k(t_{21})^{-1} k^{w}_{s,l} \\[2mm]
\kappa_{3} = k_{t,r}k_{s,r}^{-1} \quad & , & \quad
\kappa_{4} = (k^{w}_{t,l})^{-1} k(t_{41})k(t_{43})^{-1} (k^{w}_{s,r})^{-1} \ .
\end{eqnarray}
It is important to stress the two projectors in eq.\ \eqref{eq:crossingmatdef} make the supercrossing
factor independent of any gauge fixing conditions for our gauge symmetry \eqref{eq:gaugeB}. In fact
one can easily check using eq. \eqref{eq:Pprop} that any gauge transformation with some element $b$
is absorbed by the projectors.

Our main goal is to compute the matrix $\mathcal{M}$ explicitly. Note that it depends on the
insertion points $x_i$ in superspace through the dependence of the factors $k(t_{ij}) = k(t(x_{ij}))$,
that were defined in eq. \eqref{eq:factorizationij}, as well as through the factors $k_{l,r}$
in the Cartan decomposition \eqref{eq:sCartan} of the supergroup elements $g_{s,t}(x_i)$. In
order to compute the matrix $\mathcal{M}_{st}$ we  first show that it is invariant under superconformal
transformation, i.e. $\mathcal{M}_{st}(x_i^h) = \mathcal{M}_{st}(x_i)$. This then implies that it is a
function of cross ratios only and so it can be computed after moving the insertion points into
a special positions.

To see that $\mathcal{M}_{st}$ is a conformal invariant we must study the dependence of the four
tensor components one after another. We have already stated the transformation behavior of the
factors $k(t_{ij})$ at the end of the previous section, see eq.\ \eqref{eq:ktrafoh}. What we need
to study now is the transformation behavior of the factors $k_{l,r}$ in the Cartan decomposition
\eqref{eq:sCartan}. To this end let us first note that, according to eq. \eqref{eq:gijtrafoh},
the supergroup elements $g_\sigma(x_i)$ transform as
\begin{equation}
g_\sigma(x_i^h) = k(t(x_{\sigma(1)},h))\,  g_\sigma(x_i)\,  k(t(x_{\sigma(3)},h))^{-1}\ .
\end{equation}
Because of the gauge freedom of the Cartan decomposition which we described in eq.\ \eqref{eq:gaugeB},
knowing the behavior of $g_\sigma(x_i)$ under conformal transformations does not allow us to uniquely
determine the transformation law of the factors, but we can conclude that
\begin{equation}
k_{\sigma,l}(x^h_i) = k(t(x_{\sigma(1)},h)) k_{\sigma,l}(x_i) b_\sigma(x_i,h) \quad , \quad
k_{\sigma,r}(x^h_i) = b^{-1}_\sigma(x_i,h) k_{\sigma,r}(x_i) k(t(x_{\sigma(3)},h))^{-1} \
\end{equation}
for some factor $b$ that may depend on the channel, the superspace insertion points $x_i$
and the superconformal transformation $h$, yet must be the same for the left and right
factors $k_l$ and $k_r$.  For the case of $s$- and $t$-channels, these become
\begin{equation}
k_{s/t,l}(x^h_i) = k(t(x_{1},h)) k_{s/t,l} b_{s/t}(x_i,h)\quad , \quad
k_{s/t,r}(x^h_i) = b_{s/t}^{-1}(x_i,h) k_{s/t,r} k(t(x_{3},h))^{-1} \ .
\end{equation}
With these transformation laws it is now easy to verify that all four tensor components
$\kappa_i$ of the crossing factor $M$ are indeed invariant under superconformal
transformations, up to gauge transformation, i.e.
\begin{equation}
\kappa_i(x^h_k) =  b^{-1}_t(x_k,h)\,  \kappa_i(x_k)\,  b_s(x_k,h) \quad , \quad
\kappa_j(x^h_k) =  w b^{-1}_t(x_kh) w^{-1}\,  \kappa_j(x_k)\,  w b_s(x_k,h) w^{-1}\ ,
\end{equation}
where $i=1,3$ and $j=2,4$. To get the last two relations one employs the formula for
$k(t^h_{ji})$ given in eq.\ \eqref{eq:ktrafoh}). Using the definition \eqref{eq:calP}
of the projectors $\mathcal{P}_s =\mathcal{P}_t$ and the property \eqref{eq:Pprop} of
$P \in U(\mathfrak{b})$ we see that $\mathcal{M}_{st}(x_i)$ is indeed invariant under
conformal transformations.
\medskip

The analysis we have performed in this section holds for conformal and superconformal
symmetries alike. It is actually quite instructive to evaluate the final formula
\eqref{eq:crossingmatdef} for the crossing factor for spinning correlators in bosonic
conformal field theories. In this case it is in fact rather easy to obtain $\mathcal{M}_{st}$
since we can effectively reduce the problem to one on the 2-dimensional conformal group. We
will deviate from previous notations and use $G$ to denote the bosonic conformal group
$\textit{SO}(d+1,1)$ and assume $d>2$.

Since the crossing factor is conformally invariant, in computing $\mathcal{M}(u,v)$ we may assume
that $x_i$ are any points that give the correct cross ratios $u$ and $v$. In particular,
all points can be assumed to lie in the 2-dimensional plane $P$ that is spanned by the
first two unit vectors $e_1,e_2$ of the $d$-dimensional space $\mathbb{R}^d$. In this
case, the element $g_\sigma(x_i)$ is seen to belong to the conformal group of the plane,
i.e.\ $g_\sigma(x_i) \in G_P=SO(3,1)\subset G$. Within this group $g_\sigma(x_i)$ admits
a unique Cartan decomposition, which can also serve as its Cartan decomposition in $G$,
bearing in mind that the torus $A\subset G_P \subset G$ of the Cartan decomposition of
$G$ is actually a subgroup of $G_P$. Put in another way, the Cartan decomposition of
$G_P$ defines a particular gauge fixing for Cartan factors of $g(x_i)$. Note that all
relevant rotations are generated by the element $M_{12}$, which commutes with the Weyl
inversion $w$ when $d>2$. Hence we conclude that the factors $\kappa_i$ that arise
in the transition from $s$- to $t$-channel must be of the form
\begin{equation}\label{form-of-kappas}
    \kappa_i = e^{\gamma_i D} e^{\varphi_i M_{12}} \ ,
\end{equation}
for some functions $\gamma_i$ and $\varphi_i$ that depend on the insertion points $x_i$
of the four fields through their two cross ratios. Having determined the general from of
$\kappa_i$, we can find the undetermined coefficients by a direct calculation. Since we
can perform the calculation in any conformal frame we set for convenience,
\begin{align}\label{point-configurations-1}
    & x_1 = \frac{\cosh^2\frac{u_1}{2}+\cosh^2\frac{u_2}{2}}{2\cosh^2\frac{u_1}{2}\cosh^2\frac{u_2}{2}} e_1 - i
    \frac{\cosh^2\frac{u_1}{2}-\cosh^2\frac{u_2}{2}}{2\cosh^2\frac{u_1}{2}
    \cosh^2\frac{u_2}{2}} e_2\ ,\ x_2 = 0\ ,\ x_3 = e_1\ ,\ x_4 = \infty e_1\ .
\end{align}
Then it follows
\begin{equation}
    \kappa_1 = \kappa_3 = e^{\gamma D + \alpha M_{12}}\, , \quad
    \kappa_2 = \kappa_4 = e^{\gamma D - \alpha M_{12}}\ ,
\end{equation}
where
\begin{equation}
    e^{4\gamma} = \frac{x_{12}^2 x_{34}^2}{x_{14}^2 x_{23}^2}\, ,\quad
    e^{2i\alpha} = \frac{\cosh{\frac{u_1}{2}}}{\cosh{\frac{u_2}{2}}}\ .
\end{equation}
To complete this description let us also quote from \cite{Buric:2019dfk} that
\begin{eqnarray}
\label{eq:uz}
e^{u_i} & = &  1 - \frac{2}{z_i}\left(1+\sqrt{1-z_i}\right)\ , \\[2mm]
\textit{where} \quad u = z_1 z_2 & = & \frac{x_{12}^2x_{34}^2}{x_{13}^2x_{24}^2}
\quad , \quad v = (1-z_1)(1-z_2) = \frac{x_{14}^2x_{23}^2}{x_{13}^2 x_{24}^2}\ .
\end{eqnarray}
Let us note that $\mathcal{M}$ was originally defined using representations of $K = SO(1,1)
\times SO(d)$, but is computed using only representation theory of $SO(1,1)\times SO(2)$.
\medskip

\noindent
{\bf Example:} To make the last point manifest, let us give some more details for
conformal theories in $d=3$ dimensions. Let us decompose the factors $k_l = d_l r_l$
and $k_r = d_r r_r$ into dilations $d_{l/r}$ and rotations $r_{l/r}$. Following
\cite{Schomerus:2016epl,Schomerus:2017eny,Buric:2019dfk} we parametrize the elements
$r$ of the 3-dimensional rotation group through Euler angles,
\begin{equation}
    r(\phi,\theta,\psi) = e^{-\phi M_{12}} e^{-\theta M_{23}} e^{-\psi M_{12}} \ .
\end{equation}
With this choice of coordinates, the elements $\kappa_i$ have $\phi =\pm\alpha$ and
$\theta = \psi = 0$. Next let us recall that matrix elements of the spin-$j$
representation of $SU(2)$ read
\begin{equation}
t^j_{m n} (\phi,\theta,\psi) =  \langle j,m| g(\phi,\theta,\psi) | j,n\rangle =
e^{-i(m\phi+n\psi)} d^j_{m n}(\theta) \ .
\end{equation}
Here, the function $d^j_{m n}$ is known as Wigner's $d$-function. It is expressed
in terms of Jacobi polynomials $P^{(\alpha,\beta)}_n$ as
\begin{equation}
d^j_{m n}(\theta) = i^{m-n} \sqrt{\frac{(j+m)!(j-m)!}{(j+n)!(j-n)!}}
\Big(\sin\frac{\theta}{2}\Big)^{m-n} \Big(\cos\frac{\theta}{2}\Big)^{m+n}P^{(m-n,m+n)}_{j-m}(\cos\theta) \ .
\end{equation}
For $\theta=0$, the only non-zero matrix elements are those with $m=n$. Furthermore
\begin{equation}
t^j_{n n}(\pm\alpha,0,0)  = e^{\mp in\alpha}  P^{(0,2n)}_{j-n}(1) =
e^{\mp in\alpha} = \left(\frac{\cosh\frac{u_1}{2}}{\cosh\frac{u_2}{2}}\right)^{\mp\frac{n}{2}} \ .
\end{equation}
Since the stabilizer group $B = SO(d-2)$ for a bosonic conformal field theory in $d=3$
dimensions is trivial, so it the projector $P$. Putting all this together we conclude
that the crossing factor reads
\begin{equation}
    (\mathcal{M}_{st})^{ijkl}_{pqrs} = \left(\frac{u}{v}\right)^{-\frac14\sum\Delta_i}
    \left(\frac{\cosh\frac{u_1}{2}}{\cosh\frac{u_2}{2}}\right)^{\frac12(i+k-j-l)}
    \delta^i_p \delta^j_q \delta^k_r \delta^l_s\ ,
\end{equation}
where $u,v$ are the usual $s$-channel cross ratios and $u_i = u_i(u,v)$ are functions
thereof, see eq.\ \eqref{eq:uz}. The first factor in this result for the spinning crossing
factor is well known from scalar correlators. The correction it receives for spinning
correlators are diagonal in the space of polarizations but depend on the eigenvalues
of the generator $J_z$ for rotations around one particular direction $e_z$.

\subsection{Blocks and crossing symmetry equation}

In case of bosonic conformal theories, the crossing factor we have just computed
along with spinning conformal blocks is all it takes to write down crossing symmetry
constraints. For superconformal symmetries of type I, some more work is needed in order
to spell out these equations. We  describe the additional
elements in this subsection before we illustrate the entire formalism at the
example for $\mathcal{N}=2$ superconformal theories in  $d=1$ dimensions in the
next. Along the way we  also review the construction of conformal blocks from
\cite{Buric:2019rms}. In order not to clutter the presentation too much, the first
part of our discussion focuses on the $s$-channel. Other channels can be dealt with
similarly.

In Subsection 4.1 we have shown that the four-point function of primary fields in an
arbitrary representations of a conformal superalgebra of type I can we written as
\begin{equation} \label{eq:GThetaPsi}
G_4(x_i) = \Theta_{s}(x_i) \Psi_s(u_i;\theta,\bar \theta)\ ,
\end{equation}
where the supertensor factor $\Theta_{s}(x_i)$ depends on the insertion points of
the fields through
\begin{equation}  \label{eq:defOmegas}
\Theta_{s} (x_i) = \omega^{-1/2}(u_1,u_2)
\rho_{1} (k_{s,l}) \rho_{2}(k(t_{21})^{-1}k^{w}_{s,l})
\rho_{3}(k_{s,r}^{-1}) \rho_{4}(k(t_{43})^{-1} (k^{w}_{s,r})^{-1}) P_s\ ,
\end{equation}
and $\Psi_s$ is a function of the cross ratios, including all nilpotent/fermionic superconformal
invariants, that is given by
\begin{equation} \label{eq:deffs}
\Psi_s(u_i;\theta,\bar \theta) = \omega^{1/2}(u_1,u_2) F_s(\eta_{s,l}a_s \eta_{s,r})\ .
\end{equation}
In splitting eq.\ \eqref{eq:G4schannel} into a product of a supertensor factor and a
function $\Psi_s$ of the cross ratios we have included a scalar factor
\begin{equation}
\omega(u_1,u_2) = 4(-1)^{2-d}(\sinh\frac{u_1}{2} \sinh\frac{u_2}{2})^{2d-2}\coth\frac{u_1}{2}
\coth\frac{u_2}{2} |\sinh^{-2}\frac{u_1}{2}-\sinh^{-2}\frac{u_2}{2}|^{d-2}\ ,
\end{equation}
which depends on the bosonic cross $u_1,u_2$ ratios only and may in fact be interpreted as
the volume of \(K\times K\) bosonic orbits on the conformal group, see \cite{Schomerus:2017eny}
for details. The factor $\omega$ is conventional, but has some advantages that will be pointed
out below.

Let us now further analyse the factor $F_s$ in formula \eqref{eq:deffs} by expanding it in
the fermionic variables. The Grassmann variables $\theta$ that multiply the odd generators
of negative $U(1)$ R-charge in the exponent of $\eta_l$ generate an algebra $\Lambda_\theta$
while those variables $\bar\theta$ that multiply the positively charged odd generators in the
exponent of $\eta_r$ give rise to a Grassmann algebra $\Lambda_{\bar \theta}$. Before the
expansion, the wave functions $\Psi_s(u_1,u_2;\theta,\bar\theta)$ are vector valued, with
two copies of the bosonic subgroup $K$ acting in the image of $F$.
The first copy, which we refer to as $K_l$ acts on $V_{(12)} = V_1 \otimes V'_2$. Except
for the conjugation with the Weyl inversion in the second tensor component, one may think
of $V_{(12)}$ as the space of superpolarizations for the first two fields. Similarly, the
second copy $K_r$ acts on $V_{(34)} = V_3 \otimes V'_4$. When we perform the fermionic
expansion, the coefficients sit in the representation spaces
\begin{equation}
V_l = V_{(12)} \otimes \Lambda_\theta \ , \quad V_r = V_{(34)} \otimes
\Lambda_{\bar \theta} \
\end{equation}
of $K_l$ and $K_r$. Note that the bosonic subgroup $K$ acts on the two Grassmann algebras
so that indeed both spaces form a representation of $K$. We  also refer to the spaces
$V_l$ and $V_r$ as spaces of polarizations, as opposed to $V_{(12)}$ and $V_{(34)}$ which
we have called the spaces of superpolarizations.

As we explained before, the covariance properties of $F$ imply that $\Psi$ takes
values in the subspace of $B$-invariants, i.e. in the space
$$ \mathcal{T} = \left(V_{l} \otimes V_r \right)^B \ , $$
which we also refer to as the space of tensor structures. One may think of its
elements as $B$-invariant elements in the space of function of the Grassmann variables
$\theta$ and $\bar \theta$ that take values the space of superpolarizations. Let us fix
some basis of elements $\omega^I$ in $\mathcal{T}$ and denote the dual basis by $\hat{\omega}_I$.
We can collect these elements into two objects
\begin{equation} \label{eq:vdef}
v_s(x_i) = (\omega^1(x_i), \dots, \omega^T(x_i)) \, , \quad \  \hat v_s(x_i) = (\hat \omega_1(x_i),
\dots, \hat \omega_T(x_i)) \ ,
\end{equation}
where $T = \textit{dim} \mathcal{T}$ is the number of tensor structures. One may think of
$v_s$ as a rectangular matrix from the space $\mathcal{T}$ of tensor structures to the space
$V_{(12)} \otimes V_{(34)}$ with matrix elements in the Grassmann algebra $\Lambda_\theta \otimes
\Lambda_{\bar\theta} = \Lambda \mathfrak{g}_{\bar 1}$. Through the Cartan decomposition
of $g(x_i)$, the Grassmann variables $\theta$ and $\bar \theta$ are concrete functions
on the superspace $\mathcal{M}^\otimes_4$. We have  displayed this dependence on the
supercoordinates $x_i$ explicitly.

The coefficients in the fermionic expansion of $\Psi$ are functions $\psi_I$ of the two
bosonic cross ratios $u_1,u_2$ that take values in the space of $V_{(12)} \otimes V_{(34)}$
of superpolarizations. We can write this in the form
\begin{equation} \label{Psivpsi}
\Psi(u_1,u_2;\theta,\bar \theta) = v_s(x_i) \cdot \psi(u_1,u_2) = v_s(x_i) P_s
\psi(u_1,u_2) \ .
\end{equation}
Putting eqs.\ \eqref{eq:GThetaPsi} and \eqref{Psivpsi} together we can now write the
four-point function as
\begin{equation}
G_4(x_i) = \Theta_s(x_i) v_s(x_i) \cdot \psi(u_1,u_2) \ .
\end{equation}
We want to expand $G_4$ into superblocks, i.e. eigenfunctions of the super Casimir
operator. The latter turns out to take a particularly simple form when evaluated on
$\psi(u_1,u_2)$. As we have shown in \cite{Buric:2019rms}, one finds that
\begin{equation} \label{eq:sCasimir}
\textit{Cas}_s G_4(x_i) = \Theta_s(x_i) v_s(x_i)\cdot (H^{V_l,V_r}_0 + A) \psi(u_1,u_2)\ .
\end{equation}
Here $H_0$ is the spinning Calogero-Sutherland Hamiltonian for bosonic blocks, i.e.
$H_0$ takes the form
\begin{equation}
H_0 = - \frac{\partial_2}{\partial u_1^2} - \frac{\partial_2} {\partial u_2^2} +
V(u_1,u_2)\ ,
\end{equation}
where $V(u_1,u_2)$ is a potential that takes values in the space of $T\times T$
matrices. The precise form of the potential depends on the pair $(V_l,V_r)$ of
representations of $K$, but it is the same one obtains for the spinning Casimir
operator of the bosonic conformal algebra in Calogero-Sutherland gauge. In
$d=3,4$ dimensions such matrix potentials were worked out explicitly in
\cite{Schomerus:2016epl,Schomerus:2017eny}. The second term $A$ is a matrix
valued potential that was shown to be nilpotent and the precise form of these
terms is remarkably simple, see \cite{Buric:2019rms}.

The eigenfunctions of the Hamiltonian $H_0 = H_0^{V_l,V_r}$ we have just described
will be denoted by $\psi_0(\lambda_i;u_i) = \psi^{V_l,V_r}_0(\lambda_i,u_i)$. Here
$\lambda_i$ denote the eigenvalues  of the (second and higher order) Hamiltonians
which are directly related to the (spin and weight) quantum numbers of the intermediate
fields in the conformal field theory. Functions $\psi_0(u_i)$
are well studied, and explicit expressions exist at least in dimension $d \leq 4$, see in
particular \cite{Echeverri:2016dun}. Eigenfunctions of the full Hamiltonian  $\mathcal{H}$
will denoted by $\psi(\lambda_i;u_i)$. Nilpotency of $A$ guarantees that quantum mechanical
perturbation theory truncates at some order $N-1\leq\text{dim}\mathfrak{g}_+$, so that we
can obtain exact results by summing just a few orders of the perturbative expansion. It
turns out that, at any order of the expansion, the perturbation may be evaluated explicitly
with some input from the
representation theory of $SO(d+2)$.  It results in expressions superconformal blocks as
finite linear combinations of spinning bosonic blocks. In this sense our results provide
a complete solution of the Casimir equations for type I superconformal symmetry and in
particular for 4-dimensional conformal field theories with any number $\mathcal{N}$ of
supersymmetries.

Here we have described the Casimir equation and its solution for the $s$-channel but it
is clear that similar discussions apply to all channels. Reinstating the subscripts $s$
and $t$ we end up with blocks $\psi_s(\lambda_i;u^s_i)$ and $\psi_t (\lambda_i;u^t_i)$.
The eigenvalues $\lambda_i = \lambda_i(\mathcal{O})$ are related to the quantum numbers
(weight, spin, $R$ charges) of the intermediate supermultiplets $\mathcal{O}$. Let us
also stress that these blocks $\psi$ are multi-component objects with $T$  components
$\psi^I, I=1, \dots, T$ labeled by a basis of four-point tensor structures. For each
eigenvalue one can actually find $T$ independent solutions which are usually labeled
by pairs $(a,b)=ab$ of three-point tensor structures for the relevant operators
products. Consequently, the blocks $\psi = (\psi^{I,ab})$ carry two sets of labels,
an index $I$ running over four point tensor structures and an index $ab$ that enumerates
pairs of three point structures. The arguments $u_i^{s/t}$ of the blocks are functions
on superspace that are invariant under superconformal  transformations. They are related
by an exchange of the labels $2$ and $4$ and we can express one in terms of the other.
Equating the $s$- and $t$-channel expansion of the four-point function $G_4$ one finds
that
\begin{equation}\label{eq:crossing}
\sum_I\sum_{\mathcal{O}} \lambda_{12\mathcal{O},a}
 \lambda_{34\mathcal{O},b} M^{JI}_{st}(u^s_i)
\psi^{I,ab}_s(\lambda_i(\mathcal{O});u^s_i)
=
\sum_{\mathcal{O}} \lambda_{14\mathcal{O},a} \lambda_{23\mathcal{O},b}
\psi^{J,ab}_t(\lambda_i(\mathcal{O});u^t_i)\ ,
\end{equation}
where the indexes $a,b$ and $I,J$ numerate three and four-point tensor structures,
respectively, and the crossing factor $M_{st}= M_{st}(u_i)$ is given by
\begin{equation} \label{eq:crossingmatrix}
M_{st} = \hat v_t(x_i)^t \sqrt{\frac{\omega(u^t_i)}{\omega(u^s_i)}}
\mathcal{M}_{st}(u_i,\theta_s,\bar \theta_s)
v_s(x_i)\ .
\end{equation}
Note that the matrix elements of $M_{st}$ depend on the bosonic cross ratios
only. The summation in eq.\ \eqref{eq:crossing} runs over all superprimary
fields $\mathcal{O}$ in the theory. Here we have expressed the crossing factor
$\mathcal{M}_{st}(x_i)$ which we defined in eq.\  \eqref{eq:crossingmatdef}
in terms of the $s$-channel invariants and we think of the $t$-channel
invariants on the right hand side as functions of the $s$-channel ones.
Practically, it is easier to relate the $t$- and $s$-channel invariants
$u^t_i$ and $u^s_i$ to the usual bosonic cross ratios and expand on both 
sides in the nilpotent invariants. 

The additional factors $v_s$ and $v_t$ express the supercrossing factor
$\mathcal{M}$ in terms of its action in the space $\mathcal{T}$ of tensor
structures. Let us note that all three factors that appear in eq.\
\eqref{eq:crossingmatrix} are well defined and straightforward to
compute explicitly, even though the computations can be a bit cumbersome.
The only additional information one then needs in order to evaluate the
crossing symmetry constraint \eqref{eq:crossing} is the relation
between the bosonic cross ratios $u^s_i$ and $u^t_i$ in the two
different channels. These are not difficult to determine from the
Cartan decomposition. Let us stress, however, that the relations
between $u^s_i$ and $u^t_i$ involve fermionic invariants so that
there is an additional fermionic Taylor expansion to be performed
on the right hand side of eq.\ \eqref{eq:crossing} when we
express $u^t_i$ in terms of $u^s_i$. We will illustrate all this
now in the case of $\mathfrak{g} = \mathfrak{sl}(2|1)$.

\subsection{Illustration for 1-dimensional superconformal algebra}

We can now put all the above together and compute the crossing factor between the $s-$ and the $t$-
channel for the $\mathcal{N} = 2$ superconformal algebra in one dimension. To this end, the first step
is to find the group elements $g_s(x_i)$ and $g_t(x_i)$ which appear in the argument of the covariant
function $F$. In turn, this requires the supergroup elements $m(x_i)$, see eq.\ \eqref{eq:m-1d}, and
the Weyl element \eqref{eq:wmatrix}. Then one computes the products $m(x_j) m(x_i)$ and $w
m(x)$ to construct the variables $x_{ij}$ and the action of the Weyl inversion on superspace. For the
example at hand this was carried out in subsection 3.2.

These calculations provide all the input that is needed to determine $g_s(x_i)$. The supergroup elements
$g_t(x_i)$ for the $t$-channel are obtained by exchanging the labels $2$ and $4$. At this point, $g_s$
and $g_t$ depend on four sets of superspace variables, i.e. they are $3\times3$ matrices whose elements
are functions in all the $u_i,\theta_i,\bar\theta_i$ for $i=1, \dots, 4$. Since the crossing factor is
a superconformal invariant, we can apply superconformal transformations to gauge fix the coordinates of
the four insertion points. The following choice turns out to be convenient
\begin{equation} \label{eq:xigauge}
    x_1 = (x,\theta_1,\bar\theta_1),\ x_2 = (0,0,0),\ x_3 =
    (1,\theta_3,\bar\theta_3),\ x_4 = (\infty,0,0)\ .
\end{equation}
With this gauge choice, the entries of the matrices $g_s(x_i)$ and $g_t(x_i)$ depend on the bosonic
coordinate $x$ and the four Grassmann variables $\theta_{1,3}$ and $\bar \theta_{1,3}$ only.
\medskip

In the second step we have to find the Cartan decomposition for both families $g_s$ and $g_t$. For our
1-dimensional theory, the Cartan coordinates are introduced as
\begin{gather}
g=e^{\kappa R}e^{\lambda_l D}e^{\bar q Q_-+\bar s S_-}e^{\frac{u}{2}(P+K)}e^{qQ_++sS_+}e^{\lambda_r D}\ .
\end{gather}
%\begin{equation}
%    g =\begin{pmatrix}
%    e^{-\kappa+\frac12\lambda_l} & 0 & 0\\
%    0 & e^{-\kappa-\frac12\lambda_l} & 0\\
%    0 & 0 & e^{-2\kappa}
%    \end{pmatrix}  \begin{pmatrix}
%    1 & 0 & 0\\
%    0 & 1 & 0\\
%    -\bar s & -\bar q & 1
%    \end{pmatrix}    \begin{pmatrix}
%    \cosh\frac{u}{2} & \sinh\frac{u}{2} & 0 \\
 %  \sinh\frac{u}{2}  & \cosh\frac{u}{2} & 0\\
%    0 & 0 & 1
%    \end{pmatrix}     \begin{pmatrix}
%    1 & 0 & q \\
%    0 & 1 & s \\
%    0 & 0 & 1
%    \end{pmatrix} \begin{pmatrix}
%    e^{\frac12\lambda_r} & 0 & 0\\
%    0 & e^{-\frac12\lambda_r} & 0\\
%    0 & 0 & 1
%    \end{pmatrix}.
%\end{equation}
This agrees with the general prescription \eqref{eq:sCartan}, except that the torus of
elements $a$ is parametrized by a single variable $u$ in this case. Through straightforward
manipulations of supermatrices one finds the following expressions for the Cartan coordinates
of $g_s$ and $g_t$ in our gauge \eqref{eq:xigauge}. For the bosonic Cartan coordinates in
$s$-channel one has
\begin{align}\label{eq:kls}
    & \cosh^2 \frac{u_s}{2} = \frac1x \Big( 1 - \frac12\theta_3\bar\theta_3 -\frac{\theta_1\bar\theta_1}{2x} +
\frac{\theta_1\bar\theta_3}{x} +\frac{\theta_1\bar\theta_1\theta_3\bar\theta_3}{4x} \Big)\ ,\quad e^{-2\kappa_s} = 1 +
\frac{\theta_1}{x}(\bar\theta_1 - \bar\theta_3)\ ,\\[2mm]
& e^{\lambda_{s,l}-\lambda_{s,r}} = \Big(1-x-\frac12\theta_1\bar\theta_1-\frac12\theta_3\bar\theta_3+
\theta_1\bar\theta_3\Big) \Big(x-\frac12\theta_1\bar\theta_1\Big)\ ,\label{eq:lsm}\\[2mm]
& e^{\lambda_{s,l}+\lambda_{s,r}} = \Big(1+\frac12\theta_3\bar\theta_3\Big)\Big(x-\frac12\theta_1\bar\theta_1\Big)\ ,
\label{eq:lsp}
\end{align}
while in the $t$-channel these coordinates read
\begin{align}\label{eq:klt}
& \cosh^2 \frac{u_t}{2} = x\Big( 1 + \frac12\theta_3\bar\theta_3 + \frac{\theta_1\bar\theta_1}{2x}
- \theta_1\bar\theta_3 + \frac{\theta_1\bar\theta_1\theta_3\bar\theta_3}{4x}\Big)\ ,\quad
e^{-2\kappa_t} = 1 + \bar\theta_3 (\theta_3 - \theta_1)\ , \\[2mm]
&  e^{\lambda_{t,l}-\lambda_{t,r}} = - \Big(1-x-\frac12\theta_1\bar\theta_1-\frac12\theta_3\bar\theta_3+
\theta_1\bar\theta_3\Big)\Big(1+\frac12\theta_3\bar\theta_3\Big)\ ,\label{eq:ltm} \\[2mm]
& e^{\lambda_{t,l}+\lambda_{t,r}} = \Big(1-\frac12\theta_3\bar\theta_3\Big)
\Big(x+\frac12\theta_1\bar\theta_1\Big)\ .\label{eq:ltp}
\end{align}
In order to extract $\sinh(u_t/2)$ from the first and $\exp \lambda_{t,l}\ ,\ \exp\lambda_{t,r}$ from the
last two lines one has to take some square roots. Here we use the following convention
\begin{equation}
e^{\lambda_{t,r}}=i (\frac{x}{1-x})^\frac{1}{2}- \dots\ , \quad
e^{\lambda_{t,l}}=-i \sqrt{x(1-x)}- \dots \ , \quad
\sinh(u_t/2)= i\sqrt{1-x}-\dots \ .
\end{equation}
The fermionic Cartan coordinates, on the other hand, are given by the following expressions
\begin{align} \label{eq:sqs}
    & q_s = e^{\frac12\lambda_{s,r}}\Big(\theta_3 -\frac{\theta_1}{x} \Big( 1-\frac12\theta_3\bar\theta_3\Big)\Big)\ ,
      \quad s_s = e^{-\frac12\lambda_{s,r}}\frac{\theta_1}{x}\ ,\\[2mm]
    & {\bar q_s = e^{-\frac12\lambda_{s,l}}(\bar\theta_3 -\bar\theta_1)\ ,
       \quad \bar s_s = -e^{\frac12\lambda_{s,l}}\frac{\bar\theta_3}{x}}\ ,\\[2mm]
    & q_t = e^{\frac12\lambda_{t,r}}(\theta_3 - \theta_1)\ ,
       \quad s_t = -e^{-\frac12\lambda_{t,r}}\theta_1\Big(1-\frac12\theta_3\bar\theta_3\Big)\ ,\\[2mm]
    & {\bar q_t = -e^{-\frac12\lambda_{t,l}}\Big(\bar\theta_1 - \bar\theta_3\Big(x+\frac12\theta_3\bar\theta_1\Big)\Big)}\ ,
      \quad  \bar s_t = e^{\frac12\lambda_{t,l}}\bar\theta_3\ . \label{eq:sqt}
\end{align}
This concludes the second step of the construction, namely the determination of the Cartan coordinates
in the two channels.
\medskip

As a third step we want to compute the supercrossing factor $\mathcal{M}_{st}$ between the two channels
that was defined in eq.\ \eqref{eq:crossingmatdef}. Note that for our superconformal algebra the group
$K$ is generated by dilations $D$ and R-symmetry transformations $R$ only. It is abelian and hence
all its irreducible representations are 1-dimensional. Therefore, the supercrossing factor $\mathcal{M}_{st}$
consists just of a single function in the variables $x, \theta_{1,3}$ and $\bar \theta_{1,3}$. It depends,
of course, on the choice of representations for the external superfields. We shall pick four such
representations $(\Delta_i,r_i)$, corresponding to the conformal weight and the R-charges of the
superprimaries, as before. The associated representations $\rho$ of $K = \textit{SO}(1,1) \times U(1)$
were introduced in eq.\ \eqref{eq:rho-1d}. Note that in our gauge \eqref{eq:xigauge} the factors
$k(t_{41})$ and $k(t_{43})$ are trivial. Therefore, we have
\begin{align}
    & \kappa_1 = e^{(\lambda_{s,l}-\lambda_{t,l})D + (\kappa_s - \kappa_t) R}\ ,\quad
    \kappa_4 = e^{(\lambda_{t,l}+\lambda_{s,r})D-\kappa_t R},\\[2mm]
    & \kappa_3 = e^{(\lambda_{t,r}-\lambda_{s,r})D}\ ,\quad \kappa_2 =
    e^{-(\lambda_{t,r}+\lambda_{s,l}-\log x^2)D + (\kappa_s - \frac12\theta_3\bar\theta_3 +
    \frac{\theta_1\bar\theta_1}{2x})R}.
\end{align}
The matrix ${\mathcal{M}}$ can be written in terms of superspace coordinates by inserting our
explicit formulas \eqref{eq:kls}-\eqref{eq:ltp} for the Cartan coordinates in the $s$-  and
$t$-channel. This gives
\begin{align}
    \mathcal{M}_{st} & = e^{\frac{i\pi}{2}(\Delta_2+\Delta_4-\Delta_1-\Delta_3)} x^{-2\Delta_1}
    \alpha^{\frac32\Delta_1-\frac12\Delta_2-\frac12\Delta_3-\frac12\Delta_4} \times \nonumber \\[2mm]
    & \hspace*{2cm} \times \beta^{\frac12\Delta_1+\frac12\Delta_2-\frac32\Delta_3+\frac12\Delta_4}
    e^{r_1(\kappa_s-\kappa_t) +r_2(\kappa_s-\frac12\theta_3\bar\theta_3+
    \frac{\theta_1\bar\theta_1}{2x})-r_4\kappa_t},
\end{align}
where $\alpha$ and $\beta$ denote the following superspace elements
\begin{equation}\label{alpha-beta}
    \alpha = x + \frac12\theta_1\bar\theta_1,\ \beta = 1 - \frac12\theta_3 \bar\theta_3\ .
\end{equation}
In order to compute the crossing factor $M_{st}$ we are now instructed to find the map \eqref{eq:vdef}
in both $s-$ and $t$-channel. The general construction of $v$ is easy to implement since all representations
are 1-dimensional. One finds
\begin{align}
     v  = (1,q\bar q,q\bar s,s\bar q,s\bar s,qs\bar q\bar s)\ .
\end{align}
Once we insert the expressions \eqref{eq:kls}-\eqref{eq:sqt} for Cartan coordinates in the two channels
we obtain
\begin{align} \label{eq:vs-1d}
    & v_s = \Big(1,-\frac{(\bar\theta_1 - \bar\theta_3)(\theta_1-x\theta_3)}{x^{3/2}\sqrt{1-x}},
    \frac{(\theta_1 - x\theta_3)\bar\theta_3+\frac14\Omega}{x^{3/2}},\frac{(\bar\theta_1-\bar\theta_3)
    \theta_1+\frac14\Omega}{x^{3/2}},\frac{- \theta_1\bar\theta_3\sqrt{1-x}}{x^{3/2}}, \frac{\Omega}{x^2} \Big)\ ,\\[2mm]
    & v_t = \Big(1, i\frac{(\theta_1-\theta_3)(\bar\theta_1-x\bar\theta_3)}{\sqrt{1-x}},
    \frac{x\bar\theta_3(\theta_1 - \theta_3)+\frac14\Omega}{\sqrt{x}}, \frac{\theta_1 (\bar\theta_1 - x\bar\theta_3)+
    \frac14\Omega}{\sqrt{x}}, i\theta_1\bar\theta_3\sqrt{1-x}\ , \Omega\Big),
    \label{eq:vt-1d}
\end{align}
where $\Omega = \theta_1\bar\theta_1\theta_3\bar\theta_3$. Now we have all the elements that are
needed to compute the crossing factor $M_{st}$ which we defined in eq.\ \eqref{eq:crossingmatrix} as
\begin{equation} \label{eq:crossingmatrix-1d}
M_{st} = \hat v_t^T \sqrt{\frac{\sinh u_t}{\sinh u_s}}
\mathcal{M}_{st} v_s \ ,
\end{equation}
where $\sinh u$ is a special instance of the function \(\omega(u)\) that we introduced in eq.\ \eqref{eq:deffs}.
All factors that enter our expression for  $\mathcal{M}_{st}$ belong to the algebra $\mathbb{C}[x,x^{-1}]
\otimes\mathcal{A}$ where $\mathcal{A}$ is the 6-dimensional algebra that is spanned by the elements
\begin{equation}
    e_1 = 1\ ,\ e_2 = \theta_1 \bar\theta_1\ ,\ e_3 = \theta_1 \bar\theta_3\ ,\ e_4 = \theta_3 \bar\theta_1\ ,\
    e_2 = \theta_3 \bar\theta_3\ ,\ e_6=\Omega \ .
\end{equation}
If we represent the $e_i$ by the canonical (column) vector, the row vectors $v_{s/t}$ become $6\times 6$
matrices whose entries are functions of $x$. Similarly we can also turn the factor $\sqrt{\frac{\sinh u_t}
{\sinh u_s}}\mathcal{M}_{st}$ into a $6\times 6$ matrix if we replace the elements $e_i$ by their matrix
representation in the left regular representation of $\mathcal{A}$. Multiplying all these matrices the
final result is a $6 \times 6$ matrix of functions in $x$ which is given by
\begin{equation} \label{eq:crossingmatrix-1dvs2}
M_{st} = v_t^{-1} \sqrt{\frac{\sinh u_t}{\sinh u_s}}
\mathcal{M}_{st} v_s \ .
\end{equation}
Having computed the crossing factor between $s$- and $t$-channel there is only one final step left,
namely to relate the $s$- and $t$-channel cross ratios. Since the arguments of the functions $f$ in
the two channels are related by a change of variables that involves Grassmann coordinates, we need
to perform a fermionic Taylor expansion in order to write the crossing equation in terms of functions
of the bosonic cross ratio $x$ only, e.g. in the $t$-channel this expansion of \(f_t(\cosh^2\frac{u_t}{2})\)
takes the following form
\begin{equation}
    f_t = \left(1 + x\Big(\frac12\theta_3\bar\theta_3 + \frac{\theta_1\bar\theta_1}{2x} - \theta_1\bar\theta_3
    + \frac{\theta_1\bar\theta_1\theta_3\bar\theta_3}{4x}\Big)\partial + \frac14 x\Omega\partial^2\right) f_t(x)\ .
\end{equation}
Upon substitution, the crossing factor is a $6\times6$ matrix of second order differential operators
in $x$. This concludes our construction of the crossing symmetry equations for long multiplets of
$\mathcal{N}=2$ superconformal field theories in $d=1$ dimension.

\section{Conclusions and Outlook}

 In this work we have laid out a systematic theory that allows to decompose four-point functions
of local operators into superconformal blocks. It applies to all superconformal field theories
in which the superconformal symmetry is of type I, i.e.\ for which the R-symmetry contains a
$U(1)$ subgroup. This is the case for all superconformal field theories in $d=4$ dimensions and
a few other cases, in particular in $d=1,2$ and also 3-dimensional $\mathcal{N}=2$ theories.
In a first step we lifted the four point correlation function of arbitrary (long) operators to
a function on the conformal supergroup, see eq.\ \eqref{magic-formula}. A crucial ingredient in
this auxiliary step was to assign a special family of supergroup elements $g(x_i)$ to the
superspace insertion points $x_i$ of the four fields. Let us stress that this first step is
still entirely general in that it applies to all superconformal algebras and not just those of type
I. The specialization became necessary for the second step in which we introduced a special
supersymmetric version of the Cartan or KAK coordinates on the supergroup. As we had shown in
our previous work \cite{Buric:2019rms}, these coordinates are chosen to bring the Casimir
equations into a remarkably simple form that allows to construct all superblocks as finite
linear combinations of spinning bosonic blocks. The main purpose of the present work was to
determine the associated tensor factors that map functions of two cross ratios back to
the original correlation function $G_4(x_i)$ on superspace. These tensor factors consist
of two factors, a map $\Theta(x_i)$ on the space of superpolarizations, see eq.\
\eqref{eq:defOmegas}, and a map $v(x_i)$ from the space of superpolarizations to the
space of tensor structures that was defined in eq.\ \eqref{eq:vdef}. The full evaluation of
these two factors required to perform the Cartan decomposition of $g(x_i)$ explicitly. This
is in principle straightforward, but can be a bit cumbersome, in particular for higher
dimensions $d>2$. We have illustrated the explicit computation at the example of the
$\mathcal{N}=2$ superconformal algebra in $d=1$ dimension in the last subsection. Higher
dimensional examples will be treated in forthcoming work.

For some applications and in particular in order to write down crossing symmetry equations, the
tensor factors are actually not that important. What is needed, in addition to the conformal
blocks, of course, is only the ratio of tensor factors between different channels. This
quantity, which we dubbed the \textit{crossing factor} is a superconformal invariant and hence it
can be computed in any (super)conformal frame. Here we computed it for the $\mathcal{N}=2$ superconformal
algebra in $d=1$. Along with our previous results on conformal blocks for this symmetry algebra, see
\cite{Buric:2019rms}, this allows to write down crossing symmetry constraints for long multiplets,
recovering results from \cite{Cornagliotto:2017dup}. In the latter paper it was shown that the
numerical (super-)conformal bootstrap involving long multiplets is significantly more constraining
than the boostrap with the short or the superprimary components of long multiplets. Our new
derivation of these constraints, however, is now entirely algorithmic and it can be extended
without any significant additional difficulty to higher dimensional superconformal algebras of
type I. Let us also stress once again that the computation of the crossing factor in higher
dimensional theories is significantly simpler that the computation of tensor factors. We
have illustrated this at the example of bosonic conformal algebras where the computation of the
crossing factor was reduced to computations in the subgroup $\textit{SO}(1,3)$ of the $d$-dimensional
conformal group $\textit{SO}(1,d+1)$.

Our focus here was on developing the general theory. Concrete applications in particular to
4-dimensional superconformal theories will be addressed in forthcoming work. In particular
we will spell out the crossing symmetry constraint between two channels of a four-point
function involving two half-BPS and two long operators in a 4-dimensional $\mathcal{N}=1$
superconformal theory. This requires to combine all the elements of our apprroach. On the
one hand we apply the constructions and results of \cite{Buric:2019rms} to spell out the
Casimir-equations in Calogero-Sutherland gauge and we use them to construct analytic expressions
for the conformal blocks as finite sums of spinning bosonic blocks. When restricted to the
superprimary fields at the bottom of the long multiplets, our new blocks coincide with those
constructed in \cite{Li:2017ddj}. On the other hand, we evaluate our formula
\eqref{eq:crossingmatrix} for the crossing factor in the example of $\mathfrak{sl}(1|4)$.
Combining these two types of input we obtain crossing equations that can be exploited with
existing numerical techniques. Since the superblocks are finite linear combinations of
spinning bosonic blocks with coefficients whose analytical form is known, the evaluation
of the superblocks only requires the numerical evaluation of 4-dimensional spinning
bosonic blocks which has been developed in the past, see in particular \cite{Karateev:2019pvw}.
Given the experience with the long multiplet bootstrap in $d=2$ dimensions, see
\cite{Cornagliotto:2017dup}, we expect that numerical studies of the extended crossing
equations can improve on the constraints obtained from the restricted equations in
\cite{Li:2017ddj}. This may provide new clues also on the elusive minimal
$\mathcal{N}=1$ minimal superconformal field theory.

Of course it would also be interesting to spell out crossing symmetry constraints for
other correlators in superconformal theories such as the multiplets of R-currents or
the stress tensor multiplets. In principle our approach applies to such quantities as
well, as long as the superconformal algebra is of type I. Of course, applications to
various types of shorter multiplets containing conserved operators should provide
some simplifications which we did not address in this work. It would be interesting to 
study these in  more detail with a view on possible extensions of recent results in 
\cite{Manenti:2019jds}. Similarly, when summing over all operators in the crossing 
equations, one has to take shorting conditions for the blocks of short intermediate 
exchange into account. Usually it is done on the case by case basis 
\cite{Arutyunov:2002fh,Doobary:2015gia,Aprile:2017bgs,Sen:2018del}. It would be 
tempting to adopt the Calogero-Sutherland approach for a systematic analysis, at 
least in \(d=4\). Let us also mention that the situation is getting even more 
complicated in the case of non-unitary theories \cite{Yamazaki:2019yfd}. 

Another interesting direction concerns correlation functions involving non-local
operators such as boundaries, interfaces and (line, surface, $\dots$) defects
Block expansions for a large class of defect two-point functions are known, see e.g.
\cite{Liendo:2012hy,Billo:2016cpy,Lauria:2017wav,Lauria:2018klo}. A Calogero-Sutherland
theory of such blocks was developed in \cite{Isachenkov:2018pef}. It would be
interesting to supplement this by a theory of tensor structures and to extend
both ingredients, blocks and tensor structures, to the superconformal algebras.
We will return to this issue in future work. An example of physically relevant
1-dimensional defects are superconformal light-ray operators which model
high-energy scattering in supersymmetric gauge theories. Their two and
three-point correlation functions in the BFKL limit was already calculated in
\cite{Balitsky:2013npa,Balitsky:2015tca,Balitsky:2015oux}. These results may
be considered as a first step in the realisation of the bootstrap programme
for super lightray operators. Block expansions and bootstrap equations for
supersymmetric defects have also been studied and applied e.g.\ in
\cite{Liendo:2016ymz,Liendo:2018ukf,Bianchi:2018zpb,Gimenez-Grau:2019hez,
Bianchi:2019sxz}.

The restriction to type I superalgebras is certainly a limiting one that we would
like to overcome in view of possible applications of the bootstrap e.g. to the
6-dimensional $(2,0)$ theory \cite{Beem:2015aoa} or to many relevant examples of
superconformal field theories in $d=3$, see \cite{Abl:2019jhh,Alday:2020tgi,Rong:2018okz,
Atanasov:2018kqw,Agmon:2019imm} for some recent work and further references. With the
exception of the $\mathcal{N=2}$ superconformal algebra in $d=3$, non of the
superalgebras in these examples is of type I. While it is possible to treat some
special cases with methods similar to those described here, in particular in low
dimensions, it is not clear to us whether our approach does admit a systematic
extension. This remains an interesting challenge for future research.
\bigskip

\noindent
{\bf Acknowledgements:} We thank James Drummond, Aleix Gimenez-Grau, Paul Heslop, Mikhail Isachenkov,
Madalena Lemos, Pedro Liendo, Junchen Rong, Philine van Vliet  for comments and fruitful discussion. The work of ES was supported
by ERC grant 648630 IQFT. VS and IB acknowledge support by
the Deutsche Forschungsgemeinschaft (DFG, German Research Foundation) under Germany’s
Excellence Strategy – EXC 2121 ,,Quantum Universe'' – 390833306.

\appendix

\section{Proof of covariance laws}

In this appendix we prove the transformation properties of $g_{ij}$ and $k(t(y_{ji}))$
that were stated in eqs.\ \eqref{eq:gijtrafoh} and \eqref{eq:ktrafoh} at the end of section 3.
These are heavily used in section 4 in establishing conformal invariance of the crossing factor.

\vskip0.1cm {\bf Proposition} Under a superconformal transformation $h$, elements $g_{ij}$
and $k(t_{ji})$ transform as
\begin{equation}
    g_{ij} (x^h) = h g_{ij}(x) k(t(x_i,h))^{-1},\ \ k(t_{ji}^h) = w k(t(x_i,h)) w^{-1} k(t_{ji}) k(t(x_j,h))^{-1}.
\end{equation}
{\it Proof}: Consider the system of equations
\begin{align}\label{system-1}
    & m(x_i) n(y_{ji}) = g_{ij}(x),\\
    & m(x_j) k(t_{ji})^{-1} n(z_{ji})^{-1} = g_{ij}(x)w^{-1}.\label{system-1-2}
\end{align}
The first equation is the definition of $g_{ij}(x)$ and the second was shown in the course of the
derivation of eq.\ $(\ref{magic-formula})$ in section 3.3. Let us apply a transformation $h$ to all
$x_i$-s. Furthermore, we use
\begin{equation}
    m(x^h) = h m(x) k(t(x,h))^{-1} n(z(x,h))^{-1}
\end{equation}
which follows at once from definitions of $k(t(x,h))$ and $n(z(x,h))$. Doing these two steps, we get
\begin{align}\label{system-2}
    & h m(x_i) k(t(x_i,h))^{-1} n(z(x_i,h))^{-1} n(y_{ji}^h) = g_{ij}(x^h),\\
    & h m(x_j) k(t(x_j,h))^{-1} n(z(x_j,h))^{-1} k(t^h_{ji})^{-1} n(z^h_{ji})^{-1} = g_{ij}(x^h)w^{-1}.\label{system-2-2}
\end{align}
By comparing the two systems of equations we see that
\begin{align}\label{main-step}
     h^{-1}g_{ij}(x^h) = g_{ij}(x) k_{ij} n_{ij},\quad h^{-1} g_{ij}(x^h) w^{-1} = g_{ij}(x) w^{-1} k'_{ij} n'_{ij},
\end{align}
for some $k_{ij},k'_{ij},n_{ij}, n'_{ij}$. By substituting the first of these equations into the second it follows
\begin{align}
      g_{ij}(x) k_{ij} n_{ij} w^{-1} = g_{ij}(x) w^{-1} k'_{ij} n'_{ij},
\end{align}
and consequently
\begin{align}
       k_{ij} n_{ij} =  (w^{-1} k'_{ij}w) (w^{-1} n'_{ij} w).
\end{align}
However, now the grading with respect to the dilation weight implies $n_{ij} = n'_{ij}=1$. Further, a comparison
of eqs.\ $(\ref{system-1})$ and $(\ref{system-2})$ yields $k_{ij} = k(t(x_i,h))^{-1}$. Now the proposition follows from
equations $(\ref{main-step})$. Namely, the first claim is obtained by substituting the results for $k_{ij}$ and
$n_{ij}$ into the first equation. For the second, let us substitute $n'_{ij}=1$ and $k'_{ij} = w k_{ij} w^{-1}$
into the second equation in eq.\ $(\ref{main-step})$. This gives
\begin{equation}
    h^{-1} g_{ij}(x^h) = g_{ij}(x)k_{ij} = g_{ij}(x)k(t(x_i,h))^{-1}.
\end{equation}
Finally, we substitute eq.\ $(\ref{system-2-2})$ on the left hand side and eq.\ $(\ref{system-1-2})$ on the right.
Cancelling $m(x_j)$ factors, we arrive at
\begin{equation}\label{claim-2}
    k(t(x_j,h))^{-1} n(z(x_j,h))^{-1} k(t^h_{ji})^{-1} n(z^h_{ji})^{-1} w = k(t_{ji})^{-1} n(z_{ji})^{-1}
    w k(t(x_i,h))^{-1}.
\end{equation}
The grading on $\mathfrak{g}$ allows to pick the $k$-factors from both sides
\begin{equation}
     k(t(x_j,h))^{-1} k(t^h_{ji})^{-1} = k(t_{ji})^{-1} w k(t(x_i,h))^{-1} w^{-1}.
\end{equation}
Rearranging terms now gives the second claim. This completes the proof of the proposition.

\section{Supermanifolds and Lie supergroups}

In this appendix we collect some properties of supermanifolds and Lie supergroups, following \cite{Kostant:1975qe}. We hope these may be useful to some readers by offering more details on some constructions in sections 2, 3 and 4, which are however self-contained.

Recall that, by definition, a supermanifold $M$ is a topological space $X$ together with a sheaf $A$ of superalgebras, such that around any point $x\in X$ there is an open neighbourhood $U$ with $A(U)\cong C^\infty(U)\otimes\Lambda_n$, where $\Lambda_n$ is the Grassmann algebra on $n$ generators. The number $n$ is called the odd dimension of $M$. For any open set $V\subset X$, $A(V)$ is a commutative superalgebra. It is a non-trivial, but familiar, fact that the supermanifold can be completely recovered from its structure algebra $A(X)$.

Some constructions regarding supermanifolds are more easily formulated in terms of a certain coalgebra $A(X)^\ast$ rather than the structure algebra itself. The $A(X)^\ast$ is defined as the space of all elements in the full dual $A(X)'$ which vanish on some ideal of finite codimension in $A(X)$. Elements of $A(X)^\ast$ are referred to as {\it distributions with finite support}. One observes that $A(X)^\ast$ is a supercocommutative coalgebra. Namely, let $i$ and $\Delta$ be the natural injection and the diagonal map
\begin{align}
   & i : A(X)'\otimes A(X)'\xrightarrow{} (A(X)\otimes A(X))',\ \ i(v\otimes w)(f\otimes g) = (-1)^{|w| |f|} v(f) w(g),\\
   & \Delta : A(X)'\xrightarrow{}(A(X)\otimes A(X))',\ \ (\Delta v)(f\otimes g) = v (f g),\ \ v,w\in A(X)',\ f,g\in A(X).
\end{align}
Then one can show $\Delta(A(X)^\ast)\subset A(X)^\ast\otimes A(X)^\ast$, so the diagonal map makes $A(X)^\ast$ into a coalgebra. One again has that $A(X)^\ast$ determines the sheaf $A$. For example, $X$ as a set can be recovered either as the set of all homomorphisms $A(X)\rightarrow\mathbb{R}$, or as the set of all group-like elements in $A(X)^\ast$. The coalgebra $A(X)^\ast$ also plays a prominent role in the theory of Lie supergroups and their actions on supermanifolds, as will be outlined presently.

\subsection{Lie supergroups}

\vskip0.1cm Let $\mathfrak{g}$ be a Lie superalgebra, $H$ a group and $\pi : H\xrightarrow{}\text{Aut}(U(\mathfrak{g}))$ a representation of $H$ by algebra automorphisms. Further, write $F(H)$ for the group algebra of $H$. The {\it smash product} $E(H,\mathfrak{g},\pi)$ is a supercocommutative Hopf algebra constructed as follows:

\vskip0.1cm $1)$ As a vector space $E = F(H)\otimes U(\mathfrak{g})$.

$2)$ The multiplication in $F(H)$ and $U(\mathfrak{g})$ is defined in the usual way and $h x h^{-1} = \pi(h) x$.

$3)$ The comultiplication $\Delta$, counit $\eta$ and the antipode $\sigma$ are defined as
\begin{align}
    & \Delta(h) = h\otimes h,\ \Delta(x) = 1\otimes x + x\otimes 1,\ \eta(h)=1,\ \eta(x)=0,\\
    & \sigma(h) = h^{-1},\ \sigma(x) = - x,\ \sigma(A B) = (-1)^{|A||B|}\sigma(B)\sigma(A).
\end{align}
In these formulas $h\in H$, $x\in \mathfrak{g}$ and $A,B\in U(\mathfrak{g})$ are arbitrary. The set of group-like elements of $E$ is precisely $H$ and that of primitive elements is $\mathfrak{g}$. Here $\mathfrak{g}$ is identified with a subspace of $U(\mathfrak{g})$ in the obvious way. Conversely, given a supercocommutative Hopf algebra $E$ with the group of group-like elements $H$ and the Lie superalgebra of primitive elements $\mathfrak{g}$ one can show that a representation $\pi$ exists such that $E=E(H,\mathfrak{g},\pi)$.  Now assume that $\mathfrak{g} = \mathfrak{g}_{\bO}\oplus\mathfrak{g}_{\b1}$ is a Lie superalgebra and $G_0$ the connected, simply connected Lie group whose Lie algebra is $\mathfrak{g}_{\bO}$. Then there is a unique representation $\pi$ on $\mathfrak{g}$ by Lie superalgebra automorphisms which reduces to the adjoint representation on $\mathfrak{g}_{\bO}$. The smash product $E(G_0,\mathfrak{g},\pi)$ is called the simply-connected Lie-Hopf algebra associated with $\mathfrak{g}$ and denoted by $E(\mathfrak{g})$.

\vskip0.1cm A supermanifold $(X,A)$ is said to be a Lie supergroup if the coalgebra $A(X)^\ast$ is a Hopf algebra. By the above remarks, in this case $A(X)^\ast$ is a smash product $E(G_{0},\mathfrak{g},\pi)$ with $X=G_0$. In fact, if $X$ is simply connected, it can be shown that $A(X)^\ast = E(\mathfrak{g})$ for some Lie superalgebra, called the Lie superalgebra of $(X,A)$.

\subsection{Supergroup actions}

Assume now  that $G=(G_0,A)$ is a Lie supergroup and $M=(Y,B)$ another supermanifold. We will say that $G$ acts on $M$ if there is a map $A(G_0)^\ast \otimes B(Y)^\ast \xrightarrow{} B(Y)^\ast,\ u\otimes w\mapsto u\cdot w$, which satisfies
\begin{equation}
    \Delta u = \sum_i u_i'\otimes u_i'',\ \Delta w = \sum_j w_j'\otimes w_j''\implies \Delta(u\cdot w) = \sum_{i,j} (-1)^{|u_i''||w_j'|} u_i'\cdot w_j'\otimes u_i''\cdot w_j''.
\end{equation}
In this case, the structure algebra $B(Y)$ is a $A(G_0)^\ast$-module through
\begin{equation}
    \pi : A(G_0)^\ast\xrightarrow{}\text{End}(B(Y)),\ \langle w, \pi(u) f\rangle = (-1)^{|u| |w|}\langle \sigma(u)\cdot w, f\rangle.
\end{equation}
The later is called the coaction representation of $G$. The action of $G$ is fully determined by the corresponding coaction representation. Bearing in mind that $A(G_0)^\ast = E(\mathfrak{g})$, we see that a Lie supergroup action can be though of as a pair of representations of the underlying group $G_0$ and of Lie superalgebra $\mathfrak{g}$ on the vector space $B(Y)$, which satisfy a compatibility condition.

Dually, there is a map $\varphi : B(Y)\xrightarrow{}B(Y)\otimes A(G_0)$ that makes $B(Y)$ into a comodule-algebra of $A(G_0)$. This means that $\varphi$ is a morphism of algebras which is compatible with the Hopf algebra structure of $A(G_0)$. For example, $\varphi$ satisfies
\begin{equation}
    (1\otimes\Delta)\circ\varphi = (\varphi\otimes1)\circ\varphi : B(Y)\xrightarrow{}B(Y)\otimes A(G_0)\otimes A(G_0),
\end{equation}
along with a number of other compatibility conditions, see e.g. \cite{madore_1999}. Let $p$ be a point in $G_0$, considered as a morphism $p:A(G_0)\xrightarrow{}\mathbb{R}$. Then one can form the map $(1\otimes p)\circ\varphi : B(Y)\xrightarrow{}B(Y)$. For obvious reasons, we  refer to such compositions with $p$ as {\it evaluations}. Running over all points $p$, we get a representation of the $G_0$ on $B(Y)$. This agrees with the coaction representation $\pi$ from above.

In section 2.3 it was shown how the action of a superconformal group on a superspace may be computed explicitly. As was pointed out, the action requires the introduction of the tensor factor $A(G_0)=\mathcal{F}(\mathfrak{g})$ in the image of $\varphi$ - generators of $A(G_0)$ were denoted collectively by the letter $s$. In the case of bosonic transformations, $s$ could be evaluated in the sense just explained. For a fixed group element like the Weyl inversion, only the evaluated action makes sense. The evaluated action of the bosonic group $G_0$ and the infinitesimal action of the conformal Lie superalgebra described in section 2.2 fit together to form a representation of the Lie-Hopf algebra $A(G_0)^\ast$ on $B(Y) = \mathcal{M}$.

\bibliographystyle{nb}
\bibliography{bibliography}

\end{document}